\documentclass{emulateapj}
\usepackage{epstopdf}
\usepackage{graphicx}
\usepackage{natbib}
\newcommand{\beq}	{\begin{equation}}
\newcommand{\eeq}	{\end{equation}}
\newcommand{\beqa}{\begin{eqnarray}}
\newcommand{\eeqa}{\end{eqnarray}}

\def\simlt{\lower.5ex\hbox{$\; \buildrel < \over \sim \;$}}
\def\simgt{\lower.5ex\hbox{$\; \buildrel > \over \sim \;$}}

\font\tenbi=cmmib10 
\newfam\bifam  \textfont\bifam=\tenbi
\font\tenbr=cmbx10
\newfam\brfam  \textfont\brfam=\tenbr

\font\squinttenbi=cmbx10 at 9pt
\scriptfont\brfam=\squinttenbi

\def\vecnabla{
              \setbox1=\hbox{$\bigtriangledown$}
                           \raise.45ex\hbox{$\bigtriangledown$\hskip-.97\wd1
                           $\bigtriangledown$\hskip-.97\wd1
                           $\bigtriangledown$\hskip-.97\wd1}
                           \raise.47ex\hbox{$\bigtriangledown$}}

\def\rsun{\ifmmode {\rm R}_{\mathord\odot}\else $R_{\mathord\odot}$\fi}
\def\msun{\ifmmode {\rm M}_{\mathord\odot}\else $M_{\mathord\odot}$\fi}
\def\lsun{\ifmmode {\rm L}_{\mathord\odot}\else $L_{\mathord\odot}$\fi}

\newcommand{\thetaO}	{\theta_{\rm 0}}
\newcommand{\kms}	{{\rm km}\, {\rm s}^{-1}}

\def\tmb{\ifmmode {T_{\rm mb}^{13}(x,y,v)}\else $T_{\rm mb}^{13}(x,y,v)$\fi}
\def\tmbtw{\ifmmode    {T_{\rm mb}^{12}(x,y,v)} \else $T_{\rm mb}^{12}(x,y,v)$\fi}

\shorttitle{Synthetic Observations}
\shortauthors{Offner et al.}
\begin{document}

\title{Radiation-Hydrodynamic Simulations
  of Protostellar Outflows: Synthetic Observations and Data Comparisons}

\author{Stella S. R. Offner}
\affil{Harvard-Smithsonian Center for astrophysics,
    Cambridge, MA 02138}
\email{soffner@cfa.harvard.edu }
\author{Eve J. Lee}
\affil{University of Toronto, Toronto, ON, Canada M5S 3G4 }
\author{Alyssa A. Goodman}
\affil{Harvard-Smithsonian Center for astrophysics,
    Cambridge, MA 02138}
\author{H\'ector Arce}
\affil{Yale University, New Haven, CT }

\begin{abstract}

We present results from three-dimensional, self-gravitating, radiation-hydrodynamic
simulations of low-mass protostellar outflows. We construct synthetic
observations in $^{12}$CO in order to compare with observed outflows
and evaluate the effects of beam resolution and outflow orientation on
inferred outflow properties.  To facilitate the comparison, we develop a quantitative prescription for measuring
outflow opening angles. 
Using this prescription, 
we demonstrate that, in both simulations and synthetic observations, 
outflow opening angles broaden with time similarly to observed
outflows. 
However, the interaction between the 
outflowing gas and the turbulent core envelope produces significant
%bipolar asymmetry 
asymmetry between the red and blue shifted outflow lobes.
%and  variation in outflow properties. 
We find that
applying a velocity cutoff may result in outflow masses that are
underestimated by a factor 5 or more, and masses derived from
optically thick CO
emission further underpredict the mass of the high-velocity gas by a
factor of 5-10.
Derived excitation temperatures indicate that outflowing gas is
hotter than the ambient gas with temperature rising over time, which
is in
agreement with the simulation gas temperatures.
However, excitation temperatures are
otherwise not well correlated with the actual gas
temperature. 
\end{abstract}

\section{Introduction}

Young protostars power high-velocity jets that entrain
and unbind a large fraction of the natal protostellar core gas. Outflow
mass rates are
estimated to be comparable in magnitude to protostellar accretion rates \citep{bontemps96}. 
Consequently, outflows likely play an important role in
removing core mass and terminating the accretion process
\citep{matzner00, myers09}. The
largest outflows accelerate gas to velocities exceeding 100 $\kms$ and
extend across several parsecs.
%a scale that
%potentially allows for significant energy injection into the
On these scales, outflows powerfully impact their environment and
potentially inject significant energy back into the parent
molecular cloud \citep{matzner99}. This feedback may be partially
responsible 
for maintaining turbulence on ~0.1-1 pc scales
\citep{nakamura07,swift08, arce05, arce10}.
%in the cloud and
%maintaining the universally observed large molecular cloud velocity dispersions.

Due to observational resolution limits and the
inherently opaque nature of young protostellar cores, the central
outflow engine is not
well understood. Outflow morphology, velocity
distribution, and opening angle provide important constraints for
theoretical models. For example, observations suggest that outflows develop from a narrow
jet-like morphology into a wide-angle wind \citep{arce06, seale08}.
Outflow velocities also follow a Hubble-like relationship, with the
gas velocity increasing linearly from the source \citep{arce07}.
Among proposed models, the wide-angle-wind
model  \citep{li&shu96} and the
jet bow-shock model \citep{chernin&masson95} appear to best explain the observed
outflow characteristics \citep{lee00}.  

% Bachiller 90, pv, L1448
%http://adsabs.harvard.edu.ezp-prod1.hul.harvard.edu/abs/1990A%26A...231..174B
%Cernicharo, HH 111, ppv diag
%http://adsabs.harvard.edu.ezp-prod1.hul.harvard.edu/abs/1996ApJ...460L..57C
%lee00, pv diag
%http://adsabs.harvard.edu.ezp-prod1.hul.harvard.edu/abs/2000ApJ...542..925L
However, a
number of uncertainties complicate interpretation of the observations. Orientation of the
disk-outflow geometry relative to the line-of-sight make protostellar age
estimations imprecise \citep{ladd98, robit06}. 
Once the outflow gas
shocks and cools, it quickly mixes with ambient turbulent gas
and becomes impossible to distinguish. This means that
observed outflow gas may only span the previous few
thousand years of the flow. 
In addition, the ambient low-density,
high-linewidth cloud gas ($\simeq 10~\kms$) along the line-of-sight makes
identification of a
wide-angle, low-velocity outflow component problematic (e.g.,
\citealt{arce10}). 

Precession of the
jet due to rotational wobbling of the protostar may increase the
broadness and clumpiness of the outflow \citep{rosen04}. 
%Arce & GOodman 01, see Eve's reference
It is likely that variability
in the protostellar accretion rate causes variability in outflow
properties (e.g., \citealt{cernicharo96, arce02}). Consequently, bursty accretion
may produce a signature in the velocity spectrum and outflow
morphology. 
However, this is difficult to demonstrate
observationally.

%Two main models, the wide-angle-wind model and the
%jet model \citet{richer00}, appear to best explain the observed
%outflow geometries. 

In this work, we use 
gravito-radiation-hydrodynamic simulations modeling protostellar
outflows to study outflow evolution as a function of time.  
First, we introduce a quantitative method for determing opening angles
and use this to characterize the outflow properties of the raw simulation data.
Next, we investigate the
%morphological 
dependence on inclination 
%with respect to the
%line-of-sight 
and observed resolution. 
%We compare the protostellar accretion history
%SSRO putting in Hubble diagrams
%with the observed outflow properties.
%In this work we offer a quantitative method for
%measuring outflow opening angles and inclination. Our simualations
%reproduce the observed opening angle evolution in time and 
Finally, we present synthetic observations of the simulations in
$^{12}$CO. Throughout, we compare with interferometric
observations of seven protostellar outflows studied by
\cite{arce06} (henceforth AS06). These observations offer a
high-resolution picture of the inner outflow regions, the interaction
between outflow and envelope, and the evolution of outflow
characteristics over time. Since the simulations provide
complete three-dimensional information, they allow 
us to compare position-position-position (ppp) data with synthetic
and observational position-position-velocity (ppv) data and examine the
%accuracy of the 
observations.

We outline the simulation methodology and initial conditions in
\S2. We define our procedure for characterizing the outflows in \S3.
We present the data analysis, synthetic observations and observational
comparison in \S4. We summarize our results in \S5.

\section{Numerical Methods}

The simulations are performed using the ORION Adaptive Mesh
Refinement (AMR) code. For our study, we use the radiation-hydrodynamics,
self-gravitating simulations of
\citet{Offner09}, henceforth OKMK09, as initial conditions. 

The OKMK09 simulations  have a mean density of $4.46\times10^{-20}$
g cm$^{-3}$, 3D Mach number of 6.6, and a total mass of 185
$\msun$.  Energy is injected at a constant rate to maintain the level
of turbulence
(e.g. \citealt{stone98}). Particles are inserted in regions of the
flow that exceed the Jeans condition \citep{krumholz04}. 
These stars are endowed with a sub-grid stellar evolution
model that includes accretion luminosity down to the stellar
surface, Kelvin-Helmholz contraction, and nuclear burning. 
%The calculations are evolved with gravity for one cloud freefall time or 0.315 Myr.
The original calculations include the effects of radiation feedback from the
forming stars but neglect protostellar outflows, which is the subject
of our investigation. Note that we use the terms outflow and wind
interchangeably and do not distinguish between gas that is directly
launched near the protostar and gas that is swept up and entrained by this
gas.  

From the OKMK09 simulations, we select two forming protostars to
receive additional refinement beginning prior to their formation. In these high-resolution regions,  the minimum cell size is 4AU. The
remainder of the simulation is evolved with the previous number of 4
AMR levels, i.e.,  32 AU minimum cell size,  and without outflows. Although more computationally
expensive than modeling isolated protostars, this method allows us to
use self-consistent turbulent initial conditions and follow the
evolution of the outflows within a turbulent, clustered star-forming 
environment. Henceforth, we refer to the two high-resolution runs as R1 and R2.

To the selected protostars, we add a sub-grid
model for protostellar winds based upon
\citet{matzner99}. 
\citet{cunningham11} describe the details of the
model implementation in ORION, which we briefly summarize
here.
The outflow model is characterized by three dimensionless
parameters that specify the outflow ejection efficiency, outflow
velocity, and momentum
distribution. The mass ejection rate, $f_w$, gives the fraction of
infalling gas that is accelerated into a wind. This fraction is
observationally uncertain, but the disk wind 
\citep{pelletier92} and X-wind
\citep{shu94} models predict $f_w\simeq 0.1-0.33$. Here, we adopt $f_w
= 0.33$. Consequently, $1.0/(1+f_w)$ of the infalling gas accretes onto the
star, while $f_w/(1+f_w)$ is launched in an outflow.

The wind launching velocity is given by the Keplerian
velocity at the stellar surface, $v_K = \sqrt{G M_*/r_*}$. In the
case of high-mass protostars these velocities can exceed $200~\kms$, greatly
constraining the numerical timestep of the
calculation. \citet{cunningham11} solve this problem by limiting the
outflow velocity to a fraction of the Keplerian speed,
$f_v$. In the calculations we present here,  we set $f_v=1$ since 
the stars forming are low-mass with Keplerian velocities $\lesssim
100 \kms$.

%SSRO Need to switch from theta to other?
The direction of the outflow ejection is set by the direction of the
angular momentum vector of the protostar. This is determined by the
angular momentum of the accreting gas, a quantity that depends upon
the turbulent properties of the core and evolves over the calculation. 
Thus, the direction of the wind is not
fixed or set as an input parameter but is self-consistently dictated 
by the hydrodynamic evolution of the accretion flow.
As a consequence, the launching axis is roughly parallel to the angular momentum
vector of the accretion disk.

\begin{figure}
\epsscale{1.2}
\plotone{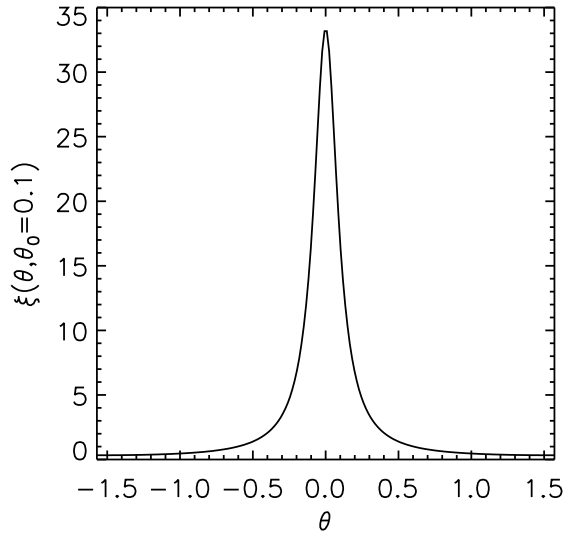}
\caption{Distribution of injected angular momentum, $\xi(\theta,
  \theta_0)$ versus polar angle, $\theta$, for $\theta_0=0.1$.}
\label{mom_dist}
\end{figure}

We define the effective opening angle of the wind, $\thetaO$,
following \citet{matzner99}. They characterize 
the angular distribution of the outflow momentum:
\begin{equation}
\xi(\theta, \thetaO)= \left[{\rm ln} \left( \frac{2}{\thetaO}
  \right)({\rm sin}^2\theta + \thetaO^2) \right]^{-1},
\end{equation} 
where $\theta$ is the polar angle measured from the protostar's
rotation axis. Through comparisons with observational data of low-mass
protostars \citet{matzner99} find that $\thetaO \lesssim 0.05$ and
suggest a fiducial value of $\thetaO=0.01$. This value results in a
strongly peaked distribution, which numerically deposits nearly all the
momentum in a couple cells along the rotational axes. Here, we adopt a
slightly larger value of $\thetaO=0.1$ as shown in Figure \ref{mom_dist}, which is comparable to the
angular width of an individual fine cell. The momentum assigned to a
cell is $\bar \xi$, the momentum distribution function averaged 
over the polar angle subtended by that cell.

The wind is injected into cells with radial distances $4 \Delta x <r
\leq 8 \Delta x $ from the protostar, so that the wind injection zone
lies outside the accretion region. This serves the
dual purpose of providing better resolution in the injection region and
allowing gas to continue to accrete onto the central protostar. In practice,
$\bar \xi$ is also set to 0 when $\theta$ becomes close to $\pi/2$.
Our outflow algorithm is fully mass conserving. Gas that was
  previously 
  automatically accreted onto the stars in the original simulation is
  instead divided between accreted gas and outflow gas that is deposited back onto the numerical grid.

%Following \citet{matzner99}, we adopt $f_{\rm wind}= 0.3$. The outflow gas is
%assumed to be atomic ($\mu =  1.28m_{\rm H}$) with a temperature of
%10,000 K.

Instead of the \citet{pollack94} dust opacity model used by OKMK09, 
we switch to an updated model from
\citet{semenov03}, which assumes
a standard iron abundance and treats the grains as composite aggregates.
The strong bow shocks
produced by outflowing gas running into ambient material can generate 
temperatures well in excess of the dust destruction
temperature. Since the outflows simulated here are young and very
low-mass, 
only a small number of cells ever reach temperatures above 1000 K. Nonetheless,
we include the treatment for atomic line cooling described by
\citet{cunningham11}, which implicitly solves for cell temperatures
exceeding 10$^4$ K.
%In this regime, atomic cooling dominates, necessitating the inclusion of an atomic cooling model. 
%We implement \citet{}  which calculates the cooling rate of the gas
%and explicitly updates the gas temperature. 

%Need to describe tff for the calculations and maybe core properties

\section{Analysis}

%Odyssey: _Outflows/idlcodes/testellip_2d_trunc.pro
\begin{figure}
\epsscale{1.2}
\plotone{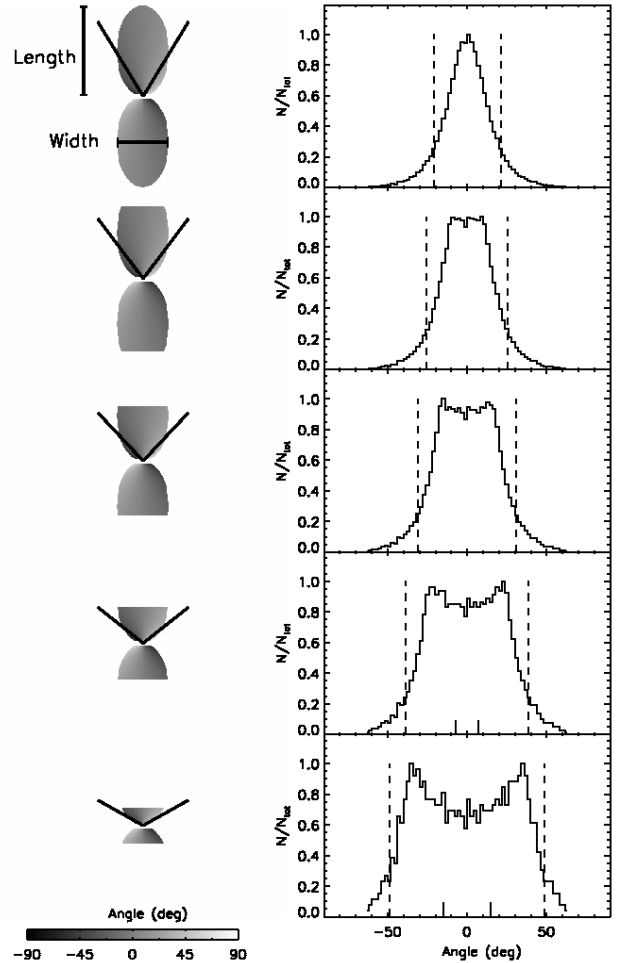}
\caption{ Left: Idealized elliptical outflow shape with half angle
  ${\rm atan}(b/a) = 20$ degrees, where $a$ and $b$ are the
   length and width of the major and minor axes, progressively
  truncated towards the center. The black lines show the
  measured opening angle. 
%The colorscale ranges from +90 (white) to
%  -90 (black). 
Right: Angle distribution of the
  pixels constituting the outflow, where the vertical lines indicate the
  opening angle defined as the full-width quarter maximum of the distribution.
\label{ellipshape} }
\end{figure}

\subsection{Outflow Identification}

Identifying the gas associated with outflowing material remains a key
observational challenge. 
Generally, outflow gas is higher
velocity, hotter, and less dense than the accreting
material. 
It is difficult to use temperature as a means to distinguish
outflow gas.
In nearby well-resolved regions, it is possible to
image outflow cavities in scattered light \citep{seale08}.
However, observers typically use low-density tracers such as
$^{12}$CO to map the outflow gas, since this data supplies both direct velocity
information and indirect estimates of the gas mass. 
%use light
%scattered off of dust grains, which traces the cavity walls and
%appears in the \textit{Spitzer} 3.6 $\mu m$ IRAC band. 
%Inferring gas
%temperatures along sight-lines is difficult \citep{shetty09}, and
%therefore 
Within molecular line data cubes, line-of-sight velocity remains the primary means of
identification. In AS06, outflows are defined by selectively
integrating over the $^{12}$CO(1-0) emission in a range of
high-velocity channels determined by visual inspection.
%Unfortunately, this process is subjective 
%and fails to include lower
%velocity gas associated with the outflow, which is obscured by
%the large turbulent linewidths of the embedding parent cloud. 

% Use projection
To identify the outflows in the simulations, we set a minimum outflow gas velocity of 2
$\kms$. 
We then generate a column
density map of cells above this cutoff, which is analogous
to an observed intensity map integrated over selected velocity
channels.

Figure \ref{ellipshape} shows the angle distributions of integrated
emission for cartoon
elliptical outflows.
In this 2D space, we treat each outflow lobe independently and
characterize outflow morphology using five  
parameters: opening angle, length, width, shape, and inclination.
To measure the opening angle, we calculate the angle with respect to the z
axis for each pixel that is above a given velocity minimum. In the
analysis, we weight all
such pixels equally.
%empirically 
The opening angle is then defined to be the full-width-quarter-maximum of the angle
distribution. This definition is empirically selected to correspond to
opening angles derived from intensity maps by eye (and protractor).
We derive the outflow inclination with respect to the vertical axis by fitting the angle distribution with either a Gaussian or higher
order polynomial. Note that an outflow may also be inclined along the
line of sight towards (or away from) the observer. However, our 
method only fits for the inclination in the plane of the sky.
For an elliptical
outflow, the inclination is the location of the distribution maximum, while for a more conically
shaped outflow, it is the local minimum. 
%SSRO We don't actually discuss the simulation outflow inclination
%Quantitative determination of the inclination is both observationally and
%numerically useful. 
Note that since the simulation inclination is obtained from the
projected distribution of cells it may not identically correspond to
the direction in which the outflow is launched. In addition,
interaction between the outflow and the envelope may affect the
inferred outflow inclination.
%

%Figure \ref{ellipshape}
%shows the angle distributions for different outflow shapes.
For a given distribution of connected high-velocity cells, the physical length of
the outflow is the longest extent measured from the center outwards
along the
outflow major axis. The width is the maximum extent
along the axis perpendicular to the outflow length. 
The Gaussian parameter, Gauss,
is the $\chi^2$ value of a Gaussian fit to the angle distribution. As
illustrated by Figure \ref{ellipshape}, the distribution of pixels for
elliptically shaped outflows is well described by a Gaussian.
Our opening angle definition
is formulated to be consistent with an elliptical outflow such that ${\rm atan}
($Width/Length$) \simeq \theta$. As we show in \S\ref{angleevol}, the pixels close to the protostar are likely to be dominated by the
beam characteristics, and thus are not suitable for an opening
angle determination.

\subsection{Opening Angle Measurement}

We first characterize an idealized, symmetric outflow geometry. Figure
\ref{ellipshape} shows an ellipse of length 0.1 pc, which has been
superimposed upon the AMR grid cell hierarchy of a simulation output. 
The effect of truncating the outflow length is apparent in 
the angle distribution evolution shown on the right. The inferred
angle increases with increasing truncation as the wide angle cells at the base of the
ellipse dominate the distribution. 

\begin{figure}
\epsscale{1.2}
\plotone{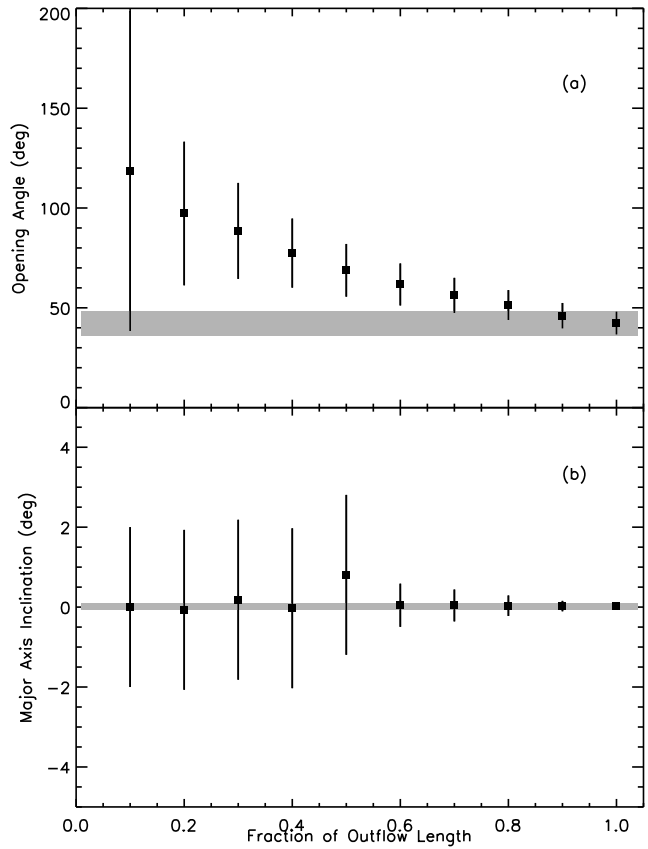}
\caption{ Top (a): Opening angle as a function of ellipse length included
  in the angle determination. (Ellipses with fractional lengths
  0.2, 0.4,
  0.6, 0.8 and 1.0 are shown in
  Figure \ref{ellipshape}). Bottom (b): Inclination of the ellipse major
  axis relative to the y axis as a function of the ellipse
  length fraction. The gray shading shows the range for the
  non-truncated case.
\label{ellipangle} }
\end{figure}

Figure \ref{ellipangle} shows the
opening angle and major axis inclination for a series of truncated
elliptical outflows. 
The error bars indicate the $\sqrt{N}$ error, where $N$ is
the number of beams contained in the ellipse assuming that the outflow
is placed 250 pc away and observed with a beam of 4''. 
Figure
\ref{ellipangle}a illustrates that the error bars do not necessarily
contain the inferred opening angle for the outflow with full
information (i.e., no truncation) even though the uncertainty increases as the
outflow length shrinks. This indicates that a poorly resolved outflow 
may be measured to have a fundamentally different opening angle than the same
outflow with complete information. 
The outflow major axis remains fixed during the truncation.
Figure \ref{ellipangle}b demonstrates that the derived
direction of the outflow axis is fairly
insensitive to the truncation.

%Odyssey: _Outflows/idlcodes/testellip_2d_trunc.pro

\begin{figure}
\epsscale{1.2}
\plotone{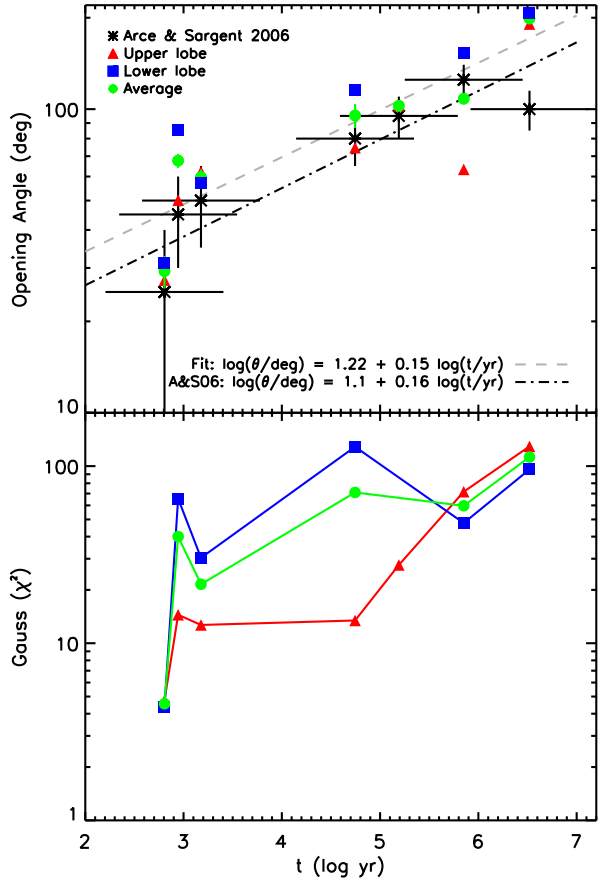}
\caption{ Opening angle (top) and Gauss parameter (bottom) for the
  \citet{arce06} sample as a function of estimated age. Stars indicate the original
values derived by \citet{arce06} with the fit shown by the
  dot-dashed line. 
The vertical error bars on the average opening angle are twice the
angle bin size used for the fit; in some cases they are smaller
  than the symbol.
%the change in the
%  measurement if the data bin size is changed by 0.1 dex. 
The dashed line shows the revised fit of the mean opening angle
derived for our angle definition.
\label{arce_prop} }
\end{figure}

For comparison, we re-analyze the AS06 data to obtain the opening
angles quantitatively. Figure \ref{arce_prop} shows the measure angles
for the upper and lower lobes and the mean of the two. The slope of
the fit we find to the data, 0.15, is quite close to the
value of 0.16 $\pm$ 0.4 found by AS06. The intercept of 1.22
is also within error of the previously published value of
1.1$\pm$0.2. This is encouraging since it suggests that our method is a
good quantitative alternative to fitting the outflow angles by
eye. One caveat to the fitting method is that for small numbers of
pixels the algorithm becomes sensitive to the bin size. We indicate
this uncertainty using vertical error bars on the plot, which are
twice the bin size used in the fit. 
%which show
%the change in the result for a bin size increased or decreased by
%10\%. 

As shown in the bottom panel of Figure \ref{arce_prop}, the angle
distribution tends to evolve from an elliptical shape (Gauss $(\chi^2)$ $\le20$)
to a more conical or irregular shape (Gauss $(\chi^2)$ $>20$). This trend may be
partially an
artifact of how the outflow is observationally sampled at late times
rather than an indication of how the geometry actually changes. This is
supported by the apparent decrease of outflow length with
time, which is most likely
because the older gas has blended with the ambient gas. 
%Or it may be rarefied, or mdot_out decreases
%Physically, 
%the maximum outflow velocity increases with time, resulting in more rarefied 
%gas with higher velocities. 
%In interferometric observations, such as those presented in AS06, the largest scales are selectively
%filtered out.

%Once the gas shocks with the ambient gas, the outflow gas inevitably
%loses its coherency and becomes indistinguishable from the background
%cloud gas.

%Analysis in 3D
%\subsubsection{Axi-symmetric Outflow Geometries}
%Analysis in 3D
%\subsubsection{Non-axisymmetric Outflow Geometries}

\begin{figure}
%\epsscale{2.1}
%\plottwo{9530_outflow_fig4-15_2.eps}{test2.eps}%{0309460_outflow_fig4-15.eps}
%\caption{ 3D volume rendering of the outflow
%  gas velocity ($\cms$) in the z direction for Run 1 at 40 kyr (top)
%  and in the x direction for Run 2 at 25 kyr (bottom). The box
%  is 0.04 pc on a side and the data is plotted with fixed cell resolution
%  of $dx=65$ AU.
\epsscale{2.3}
%yt_outflow_image_R1.py, yt_outflow_image_R2.py
%{pltgbw09730_0_0_cp333.eps}{pltgbbw309460_2_0_cp333.eps}
\plottwo{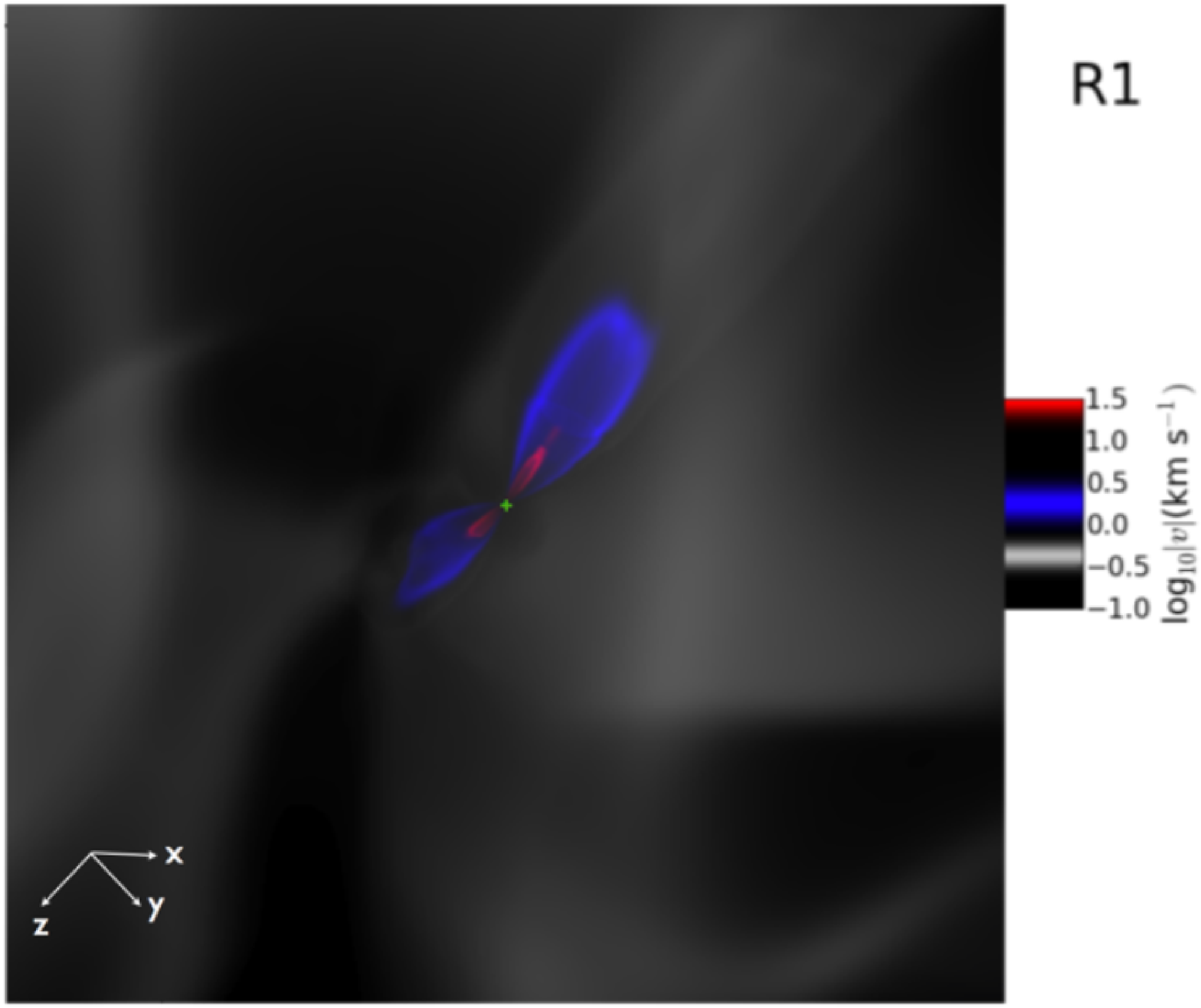}{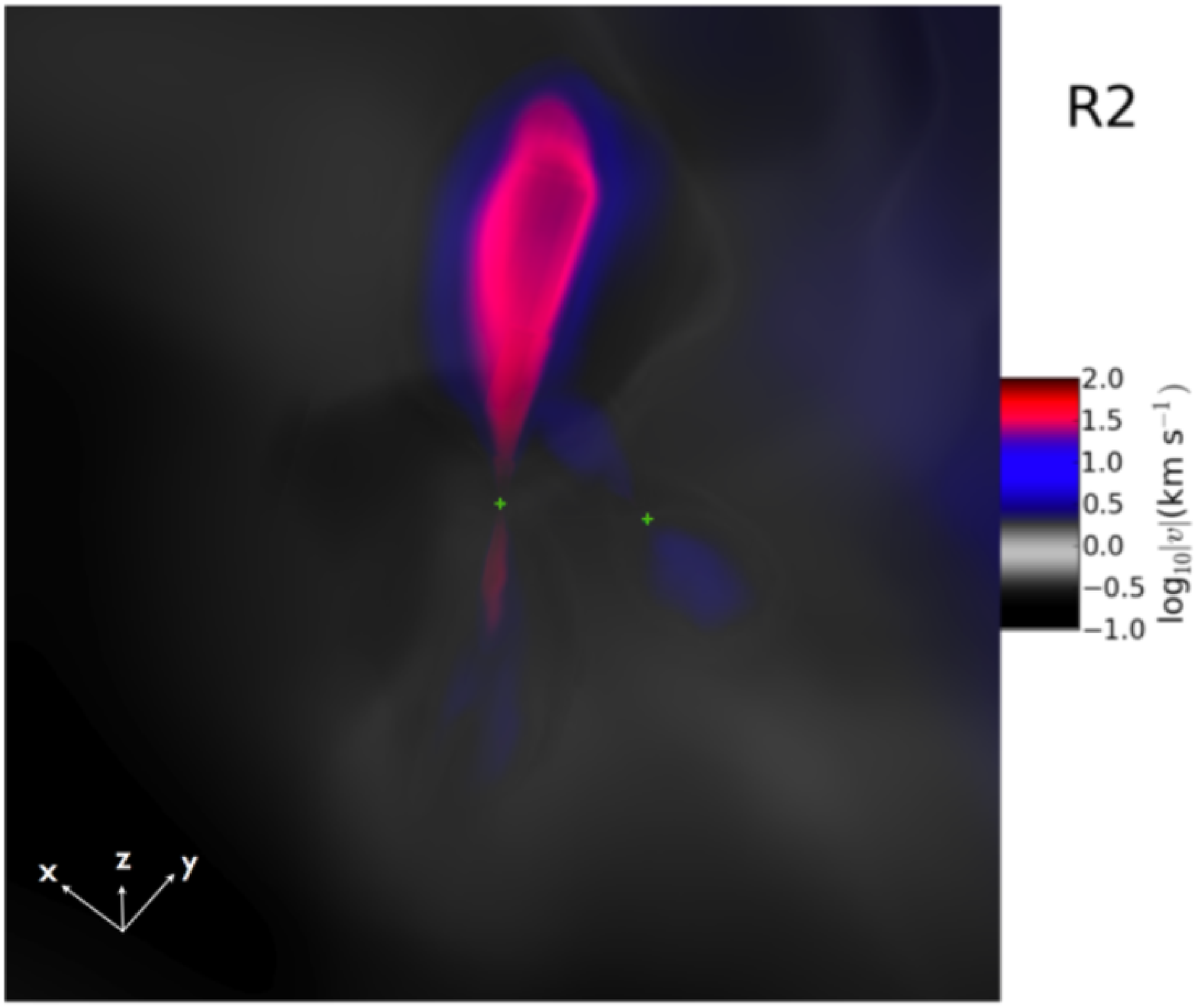}
\caption{ Volume rendering of the outflow
   gas velocity (km s$^{-1}$) for R1 at 40 kyr (top)
and for R2 at 20 kyr (bottom). The box
  is 0.15 pc on a side with the stellar positions marked by green crosses. 
\label{3dviz} }
\end{figure}

\section{Results}

\subsection{Outflow Morphology}

Figure \ref{3dviz} shows a volume rending of the simulation
velocities for both runs. The outflow axis in R1 is almost
directly aligned with the z axis so that $v_z$ traces the fastest
moving gas, which ranges up to 70 $\kms$. R2 forms two protostars
with fairly distinct outflows. The secondary protostar arises from
fragmentation in the core rather than disk fragmentation, so that the outflow
axes of the two are somewhat misaligned. However, the two R2
protostars are sufficiently close that separating the outflow lobes in 
projection is difficult. Consequently, in the following analysis we will use R1 to
explore outflow evolution for individual sources, while R2 will illustrate a
case in which observations are confused by the presence of a 
second, unidentified outflow source.

All three outflows shown in Figure \ref{3dviz} exhibit a significant amount
of asymmetry between the lobes. The sub-grid model launches the
outflow gas from the grid symmetrically about the protostars. (On the
scale of the figure the outflow launching region is contained in the
central pixel.) This means that any asymmetry that arises is directly
the result of the interaction between the outflow and the turbulent
envelope. For example, the R1 protostar forms in a more filamentary
structure and the net angular momentum direction is almost directly
aligned with the filament axes. As a result, the positive $z$ outflow
lobe is diverted from the launching axis and confined by the dense
filament gas. The outflow lobe along the negative $z$ axis breaks out of the filament and
extends further. In R2 asymmetry in the core also leads to
mismatched outflow lobe sizes.

%\section{Synthetic Observations in CO}

\subsection{Opening Angle Evolution}\label{angleevol}

%SSRO Maybe use a larger change to make the lower error bars larger) 
%SSRO Note that the outflow may be longer, but the length indicates
%the size of the connect region with vel >= 2km/2 at the star
%Odyssey: _Outflows/idlcodes/plt_hfang.pro
%I think the length scaling is off by a factor of 2
\begin{figure*}
\epsscale{1.18}
\plotone{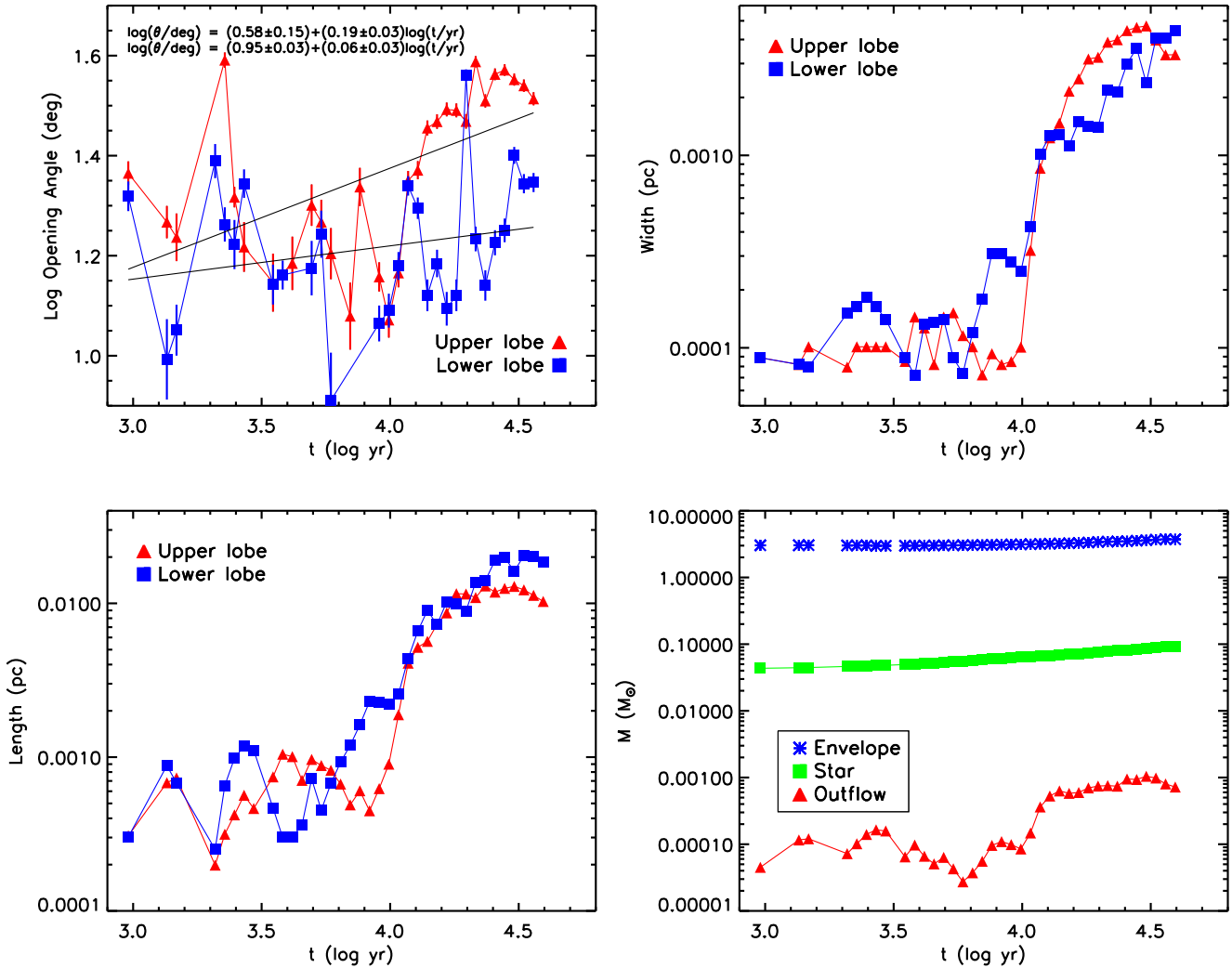}
\caption{ Opening angles, outflow length, outflow width, and mass  for the simulation as a function of time,
  where the angle are calculated for the integrated column of the cells
  belonging to the outflow as defined by a velocity $\ge 2 \kms$.
The simulation data has been divided into
  45 logarithmically spaced bins.
  The vertical error bars indicate the error given the number of beams
  contained in the outflow if it is 250 pc away and observed with a
  4'' beam. The core mass is defined as the sum over cells with densities
  $> 10^4$ g cm$^{-3}$.
%(Points with no errorbars indicate that the outflow area is $\lt$
 % 2 beams.)
\label{sim_prop} }
\end{figure*}

Figure \ref{sim_prop} shows the R1 outflow evolution as a function of time.
The outflow gas is identified using a velocity cutoff of 2 $\kms$, where
the full 3D information is used to determine
the outflow and envelope masses. These cells are projected along the
$x$-direction so that the outflow opening angle, width, and length is
calculated when the outflow axis is nearly parallel to the $z$-axis. A
fit to the opening angles demonstrates that the lobes widen with time 
similarly to observations. 
The earliest time only contains a small number of cells
in the outflow so that it is artificially broadened, but a strong trend
is apparent for $3.7 \leq {\rm log}(t)\leq 4.7$. 
(See Section 5 for discussion of the origin of this
broadening.) However, the different outflow lobes in Figure
\ref{sim_prop} differ from one another and display significant
shorter-time variation. For example, the upper outflow lobe growth
stalls at late times. This suggests that even if outflows generally
evolve with time, variation in individual outflow behavior can be large.

Considering the large
uncertainties in observed outflow ages, poorly constrained
line-of-sight inclinations, and instrument limitations, the agreement between
the simulations and the observations shown in Figure \ref{arce_prop}
is striking. This supports the finding of AS06 that 
outflow angles evolve with time. However, the protostellar masses in
these calculations are $\lesssim 0.1~\msun$ for the time of
comparison, so they are still very young. Outflow sizes are
$\lesssim 0.1$pc, although this is similar to the sizes of those in
AS06, which are $\sim 0.005-0.1$pc. Figure \ref{sim_prop} illustrates the
length of only the connected cells with velocities $\ge 2 \kms$. These
cells are also very low-density and have a total mass of $\sim
0.001~\msun$. 

%These are both in the 
\begin{figure}
\epsscale{1.2}
\plotone{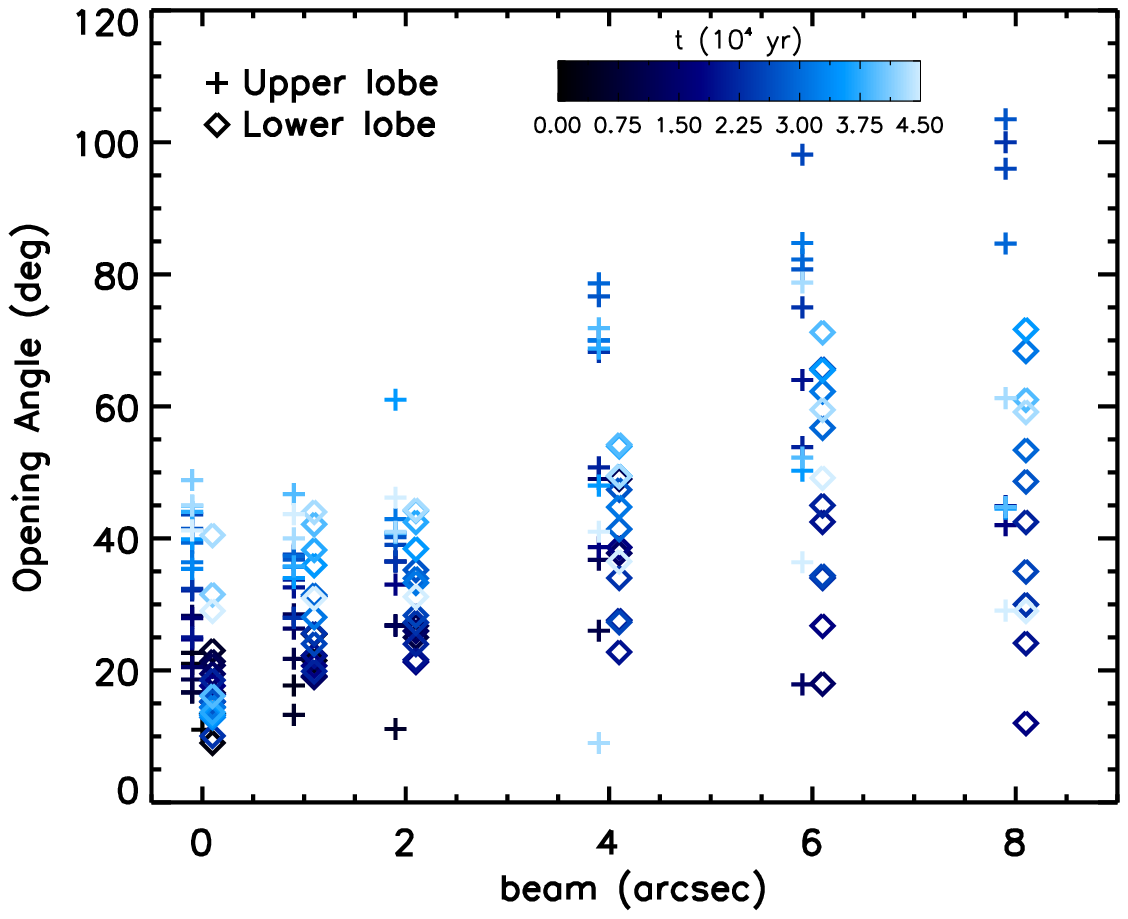}
\caption{ Opening angle versus beamsize  assuming that the outflow is
  observed at a distance of 250 pc. The color scale indicates the age
  of the outflow in units of $10^4$ years. Symbols for the different
  lobes have been offset for clarity.
%(Points with no errorbars indicate that the outflow area is $\lt$
 % 2 beams.)
\label{ang_vs_beam} }
\end{figure}

The R1 opening angle extrapolated at $t=0$ is lower than that found by AS06, although the positive
$z$ fit is still within the observational error. Smearing due to the
observational beam likely accounts for some of this
discrepancy. Figure \ref{ang_vs_beam} shows the measured opening angles
for R1 as a function of convolution with different beam sizes. For a
4'' beam, which is comparable to the resolution of AS06, some angles may be
broadened by up to $\sim$80\%. Broadening becomes more significant for
larger beam sizes. The earliest times in the simulation, 
for which the outflow is already
broadened by the small number of cells, appear least affected by the beam
smearing.

\subsection{Synthetic Observations}

While comparison between the simulation and observed $^{12}$CO data in \S\ref{angleevol}
is suggestive, it is important to compare the data directly with synthetic observations 
in $^{12}$CO. In this section we use the radiative transfer code
MOLLIE \citep{keto10} to model the emission from the first seven
$^{12}$CO and $^{13}$CO transitions. We assume a standard abundance of $5.6\times 10^{-5}$ $^{12}$CO
relative to H$_2$ and an abundance of $7.3\times 10^{-5}$ $^{13}$CO
relative to H$_2$, which are comparable to CO estimated abundances in
low-mass envelopes and outflows \citep{carolan08, carolan09}. However,
there are a wide range of inferred CO
abundances among different cores and star-forming regions, values of
which are often individually uncertain by a factor of two or
more. These uncertainties complicate the estimation of gas mass beyond
the already challenging problem of identifying the outflow gas and 
accounting for projection. 
%include the effects of depletion onto dust
%grains. 

We perform the radiative transfer calculation on a cube of side
0.15 pc centered on the protostar. We regrid the AMR data to 
produce a grid  of 128$^3$ with 130 AU cell resolution nested within a
grid of 128$^3$ and 260 AU cell resolution.
%We then XCO = 1.8$\times 10^{20}$
%cm$^{-2}$K$^{-1} \kms$ \citep{pineda08} to convert the
%integrated CO intensity to mass.

\begin{figure}
\epsscale{1.2}
\plotone{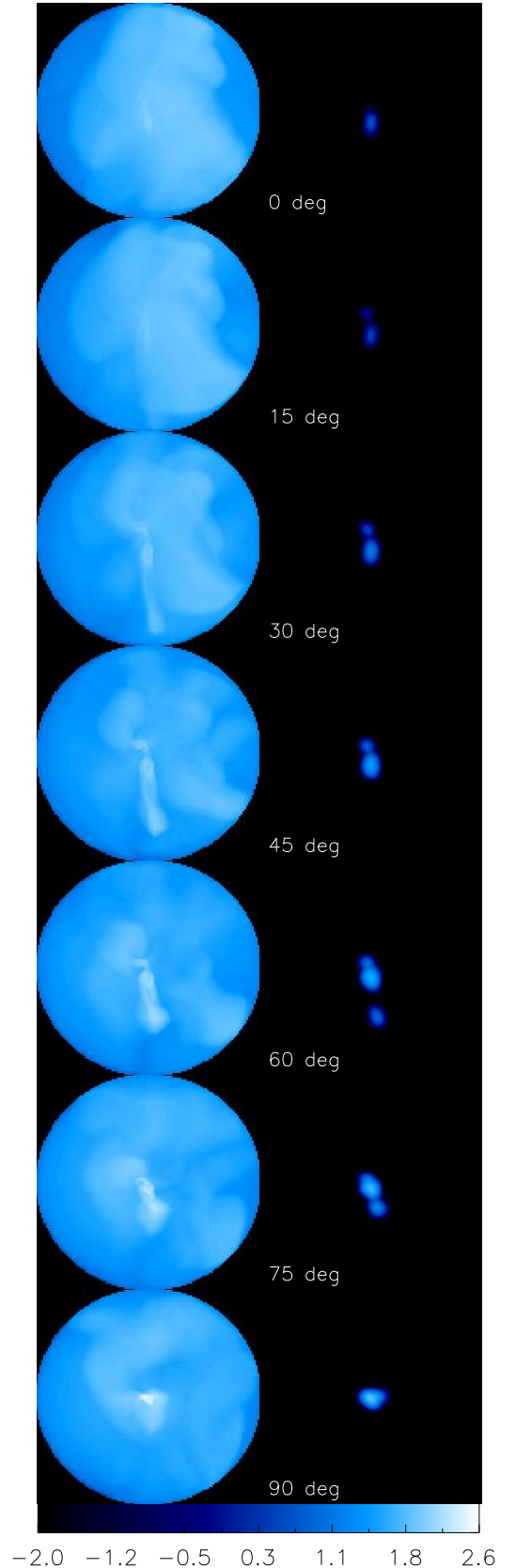}
%\plotone{pltgbw07330_Outflows_beam_4_7_fig.eps}
\caption{ Log CO intensity (K) for R1 at $t=48$kyr. Inclination with respect to the line-of-sight
  decreases vertically. The diameter of each map is 0.15 pc.
  The left column shows maps including all gas, while the right column shows
  integrated mass only for the outflow gas ($|v-v_*|\ge 2 \kms$). On the
  right, the data has been convolved with a 4'' beam assuming that the
  outflow is located at 250 pc. 
\label{COinten} }
\end{figure}

Figure \ref{COinten} shows integrated $^{12}$CO(1-0) emission maps from 
observations of R1 at different inclinations with respect to the
line-of-sight. This includes all gas with velocities in the range $\pm
20~\kms$, which encompasses the minimum and maximum outflow velocity
at 130 AU resolution.\footnote{In the central launching region 
in which the AMR resolution is 4
AU, velocities reach $\sim$ 100 $\kms$. These velocities occur in a
small volume and are reduced by the flattening to constant pixel size.}
The outflow structure is clearly visible in the integrated
maps. Such structure may not be apparent in observed integrated
maps due to emission from ambient cloud gas between the source and
the observer. The right panels show the map an observer would make
when integrating over gas in higher velocity channels ($|v| \ge 2$ km s$^{-1}$).

%SSRO  reference to seale & looney?
The lower outflow lobe shows substructure that could be
interpreted as the result of accretion bursts. However, the accretion history,
and hence the outflow mass ejection rate,
is fairly smooth (see Figure \ref{synth_sim_mass}). Instead, the 
discontinuous morphology results from
the outflow interacting with the turbulent and asymmetric core gas.

%Odyssey: _Outflows/idlcodes/plot_outflow_fig.pro
%Change this to be in K

\begin{figure}
\epsscale{1.2}
\plotone{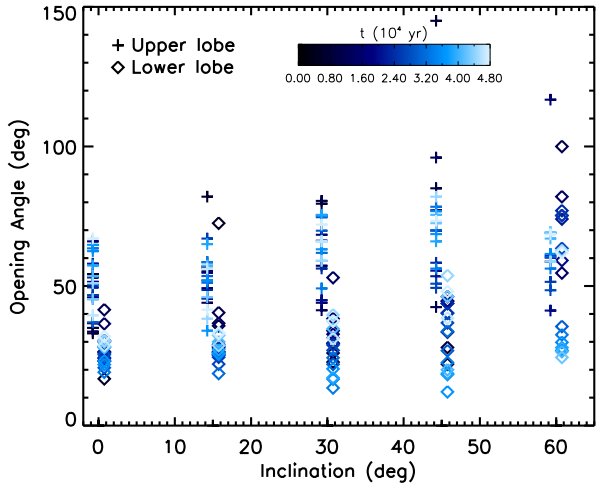}
\caption{ Opening angle versus inclination (see Figure 7) assuming that the outflow is
  observed in $^{12}$CO at a distance of 250 pc. The color scale indicates the age
  of the outflow in units of $10^4$ years. Symbols for the different
  lobes have been offset for clarity.
%(Points with no errorbars indicate that the outflow area is $\lt$
 % 2 beams.)
\label{ang_vs_incl} }
\end{figure}

Figure \ref{ang_vs_incl} shows the opening angle versus
inclination measured from the $^{12}$CO emission maps with no beam smearing. The opening angle appears slightly
broader as the outflow axis becomes more parallel to the line-of-sight, but
there is not a strong dependence on the inclination. At
low inclinations, the highest velocities are mainly perpendicular to the
line-of-sight and the angle is measured using only a small sample of
cells. 
When the outflow motion is mostly
along the line-of-sight, the lobes become shorter and rounder.
 Observationally,  outflow orientations with completely overlapping lobes
are not well suited to measuring opening angles compared to 
those with intermediate inclinations. 
Although the outflow inclination is difficult to infer from
projection, the axes of the outflows
investigated by AS06 are likely inclined between 30 to 60 degrees with
respect to the line-of-sight. 
%We thus perform the analysis when the
%outflow are tilted at 45 degrees.

%_Outlfows/idlscripts/plot_hfang_flatten.pro, plot_view_fig
%These are both in the 

%Odyssey: _Outflows/idlcodes/plot_views.pro
%Odyssey: _Outflows/idlcodes/plot_view_fig.pro based on plot_views
%\begin{figure}
%\epsscale{1.0}
%\plottwo{pltgbw06730_9730_hfang_1_OpenAngle_v2.eps}{pltgbw06730_9730_hfang_1_Incl_v2.eps}
%\caption{ Opening angle (top) and plane-of-sky inclination (bottom) of
%  a simulated outflow at X kyr (thin) and 50 kyr (thick) observed in $^{12}$CO with 1'' beam as a function of
%  tilt with respect to the observer line-of-sight.)
%\label{tilt} }
%\end{figure}

%Odyssey: _Outflows/idlcodes/plot_angle_time_evol.pro
\begin{figure}
\epsscale{1.15}
%Converted from Alyssa's pdf using:
%pdftops Opening_ang_vs_t_Alyssa.pdf -eps
\plotone{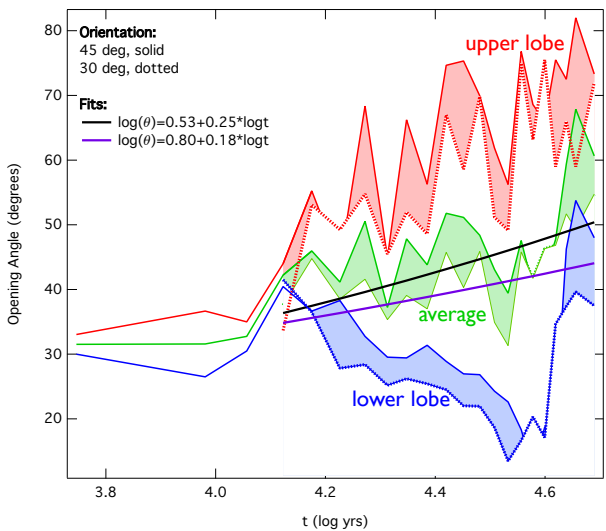}
%\plotone{Angle_vs_logt1_lng_0_lat_45_vshift_v212vw.eps}%Angle_vs_t_1_lng_0_lat_45_vshift_v212vw.eps}
\caption{ R1 opening angles as a function of time,
  where the angle is calculated using the integrated $^{12}$CO(1-0)
  intensity map including emission from channels 
  with $|v|\ge 2~\kms$ and the beamsize is 1''. The
  thick and thin lines indicate 45 and 30 degree inclinations,
  respectively, 
relative to the line-of-sight.
\label{synth_sim_prop} }
\end{figure}

Figure \ref{synth_sim_prop} shows the opening angle versus time using
the CO intensity maps to calculate the angle. Like Figure
\ref{sim_prop}, the angle increases on average as a function of time.
Linear fits to the average opening angles in Figure
\ref{synth_sim_prop} gives $\theta = 0.80 + 0.18~{\rm log}(t)$ and $\theta =
0.53 + 0.25~{\rm log}(t)$ for inclinations of 30 and 45 degrees,
respectively.  
However, like Figure \ref{sim_prop}, the two lobes show
different trends with time. For the projected data, the lower
outflow lobe is nearly flat as a function of time, while the upper
lobe widens significantly. This is mainly due to the concurrent widening
and elongation of the lower lobe.
Despite the variation, this suggests that
trends inferred from molecular ppv information are
reflective of the actual ppp information. In this case, the similarity is
particularly strong because CO is an efficient tracer of the low-density
gas within the outflow cavity.

%If the data is instead observed with 4'' beam
%resolution and the earliest opening angles are included (which are
%artifically broadened) then the opening angles \textit{decrease} in
%time with: $\theta = xx + xx log(t)$ and $\theta =
%xx + xx log(t)$. Positive $d\theta /dt$ is recovered if the fit is
%performed over $t \ge 10$ kyr.

\subsection{Outflow Mass Evolution}

%What is the temperature for H2, L1448 citation?
Molecular hydrogen has no dipole
moment and is observationally invisible unless the gas is strongly shocked
\citep{yu99}. Consequently, CO, which is the next most abundant
molecular species, serves as the primary tracer
of molecular gas. In this section we consider two methods for
estimating molecular gas mass from CO emission.

For an assumed abundance of CO, a conversion factor can be used to
relate CO intensity to total gas mass. This quantity, known as the $X$-factor, is defined as
\begin{equation}
 X_{\rm CO} = \frac{N(H_2)}{W(^{12}CO)}, 
\end{equation}
where $W(^{12}$CO) is the $^{12}$CO(1-0) integrated intensity and $N(H_2)$ is
the column density of molecular hydrogen.  
%This factor is particularly crucial
%for studying distant molecular clouds and extragalactic star
%formation. 
%since other molecular gas tracers and dust emission are either
%not well resolved or not detectable.  
%Using the $X$-factor to determine the gas mass is complicated since
The abundance of CO depends upon the local gas density,
temperature, and UV field, so that in reality, the $X$-factor 
varies widely over an individual cloud \citep{pineda08,
  glover11}
%, but it is generally correlated with total
%gas mass. 
However, this complicated chemistry is usually reduced to
an average $X$-factor.  As discussed in section 4.3, we adopt a typical mean CO
abundance from observations.  Here, we use the simulated outflows
to assess how accurately the standard $X$-factor and other more
complicated methods recover  
outflow mass. 

We adopt $X_{\rm CO} = 1.8\times
10^{20}$cm$^{-2}$K$^{-1} \kms$, which is the mean value for molecular clouds in
the solar neighborhood \citep{dame01}. \citet{pineda08} found that this
factor overstimated the gas mass by a factor of
45\% compared to the estimated extinction
mass. However, the typical mean abundances inferred by \citet{pineda08} are
a factor of $\sim$10-30 higher than the value we assume for our molecular
line transfer calculation, where the difference is due to lower expected
abundances for dense core and outflow gas versus lower density cloud
gas (e.g., \citealt{carolan08}).
 
\citet{pineda08} found that the $X$-factor is most reliable when
the gas is both optically thin and $A_v > 4$. While the emission from the
outflow gas alone is marginally optically thin (with increasing
optical depth as the inclination approaches 90 degrees and 
lower velocities), the integrated $^{12}$CO(1-0) emission
through the core envelope is optically thick with $\tau > 10$. This
means that the emission is saturated and $X_{\rm CO}$, which assumes a linear relationship between emission and gas mass, will
underestimate the outflow gas mass.
%, particularly along sightlines where the
%highest velocities are along the line of sight.

\citet{bally99} and \citet{yu99} present a more nuanced approach for obtaining the outflow gas mass from CO
emission. 
Their
methods are similar to that of \citet{arce01}, henceforth AG01, which
we follow here. All three techniques improve upon
the estimation above by correcting for optical depth effects and
including more of the
low-velocity outflow gas mass. We outline the AG01 proceedure below and
post-process the simulations accordingly 
(see AG01 and references therein). 

%Need to state that run 13CO data cubes
While the $^{12}$CO(1-0) emission may be quite optically thick, it is possible
to utilize the more optically thin $^{13}$CO(1-0) emission to calculate the
true optical depth and correct the $^{12}$CO(1-0) mass estimate. 
In the first step of the procedure, we spatially average over all the $^{12}$CO(1-0) and
$^{13}$CO(1-0) spectra and derive the ratio of the two lines, $R_{12/13}$, as a function
of velocity. The resultant parabolic line ratio shape  indicates that
the opacity is not constant with velocity as many studies assume. 

We then  perform a polynomial fit of $R_{12/13}(v)$, which is constrained to reach a minimum at the
cloud velocity. Following AG01, who limit the velocity range
to exclude a secondary coincident cloud, we limit the fitted velocity range
to within $1.5~\kms$ of the minimum. Using this fit, we extrapolate to 
the high-velocity wings, where observationally $^{13}$CO is too weak to be
detected. (Even in the noiseless synthetic observations, the $^{13}$CO
emission is negligable at velocities above a few $\kms$.)

We next derive an effective ``$^{13}$CO main beam temperature,'' \tmb, to 
estimate the $^{13}$CO opacity. Observationally, if the $^{13}$CO(1-0) emission is greater than
twice the rms noise, then \tmb~  can be used directly. Otherwise, the
$^{13}$CO main 
beam temperature must be inferred from $^{12}$CO: 
$\tmb=\tmbtw/R_{12/13}(v_i)$. 
We adopt a rms noise value of 0.06 K per 0.2 $\kms$ channel, similar
to that used by AG01. 

If $^{13}$CO(1-0) is indeed optically thin, then its opacity may be estimated:
\begin{equation}
\tau_{13}(x,y,v) = - {\rm ln} \left( 1- \frac{\tmb}{T_0/[{\rm exp}(T_0/T_{\rm
	ex})-1]-0.87}  \right),
\end{equation}
where $T_0 = h \nu /k =5.29$ and $T_{\rm ex}$ is the excitation
temperature under the assumption that $^{12}$CO(1-0) is optically thick:
\begin{equation}
T_{\rm ex} = \frac{5.53}{\rm{ln}[1+5.53/(T_{\rm peak}+0.82)]}.
\end{equation}
Here, $T_{\rm peak}$ is the peak temperature for
the $^{12}$CO emission. We find $T_{\rm ex}\sim$10-12 K, in good agreement
with the known simulation ambient gas temperature of 10 K.

Finally, we derive the column density of $^{13}$CO for each pixel:
\begin{equation}
N_{13}(x,y,v) = 2.42 \times 10^{14}(T_{\rm ex} + 0.88) \frac{\tau_{13}(x,y,v)
  dv}{1-{\rm exp}(-T_0/T_{\rm ex})},
\end{equation}
where $dv$ is the channel width in $\kms$.\footnote{Alternatively,
  for optically thin gas the column density can be expressed as 
\begin{equation}
N_{13}(x,y,v) = \frac{2.42 \times 10^{14}(T_{\rm ex} + 0.88)\tmb
  dv}{[1-{\rm exp}(-T_0/T_{\rm ex})]\times[J(T_{\rm ex})-J(T_{\rm
  bg})]}, \nonumber
\end{equation}
where $J(T) = T_0/[{\rm exp}(T_0/T)-1]$ and $T_{\rm bg} = 2.73$ K is the
  background temperature. We find that this lowers the derived mass
  estimate by $\sim$10\%.}
The total mass per pixel of area, $A$, is given by
\begin{equation}
M(x,y,v) = m_{\rm H_2}N_{\rm H_2}(x,y,v)A, 
\end{equation}
where $m_{\rm H_2} = 2.72 m_{\rm H}$ is the mean molecular weight and
the conversion between the $^{13}$CO column density and the molecular
hydrogen column density is assumed to be $N_{\rm H_2} = 1.4 \times 10^6
N_{13}$. Note that the exact value depends upon the local $X$-factor. When modeling the HH300 outflow in the Taurus star forming
region, 
AG01 adopt $N_{\rm H_2} = 7 \times 10^5 N_{13}$, a generic value for
dense gas in Taurus that was obtained by \citet{frerking82}.
%What is this actually in the simualtions?

\begin{figure}
\epsscale{2}
%/n/itc1/soffner/_gbwplots/_2dproj_flatten_obs/
\plottwo{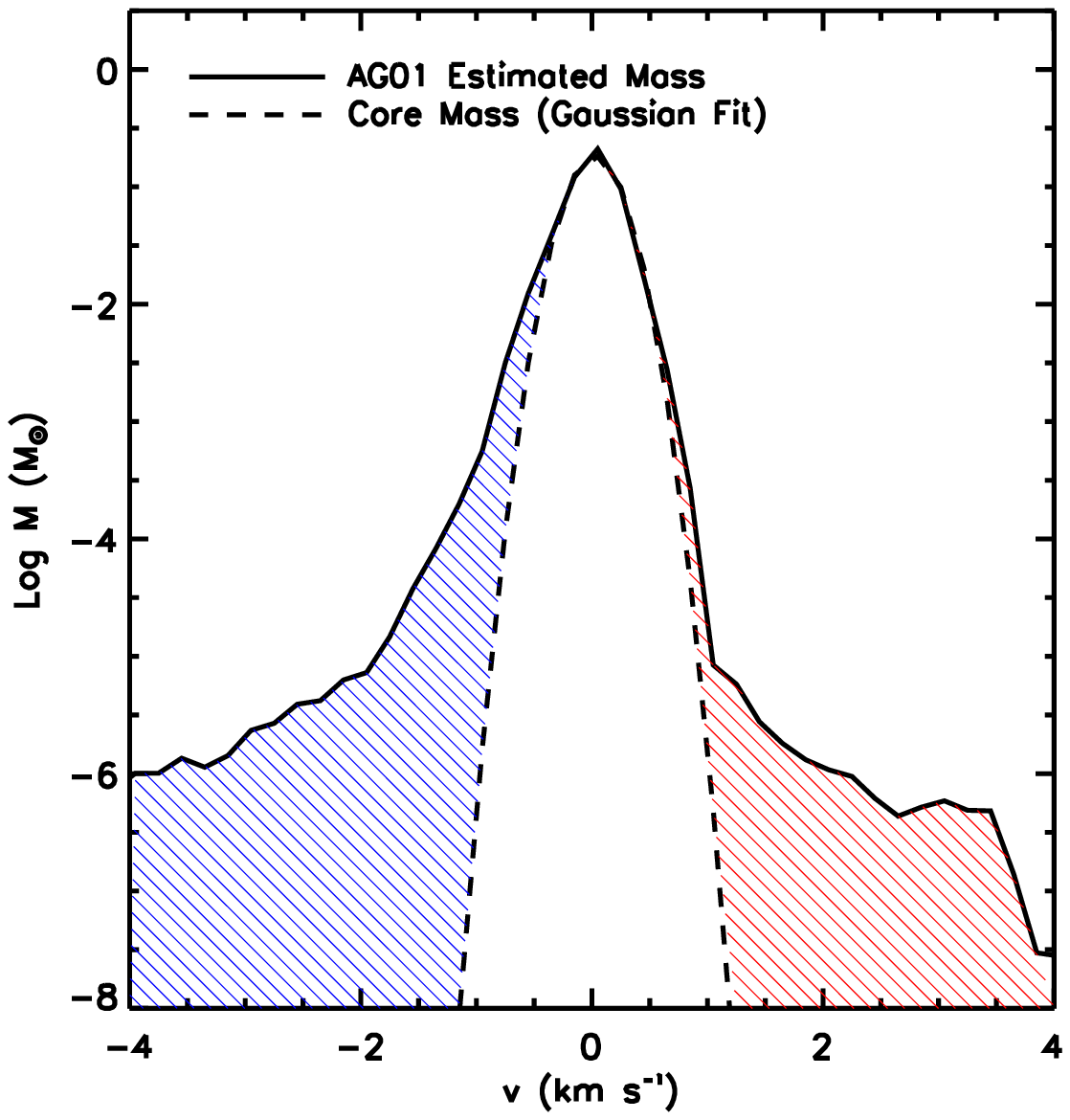}{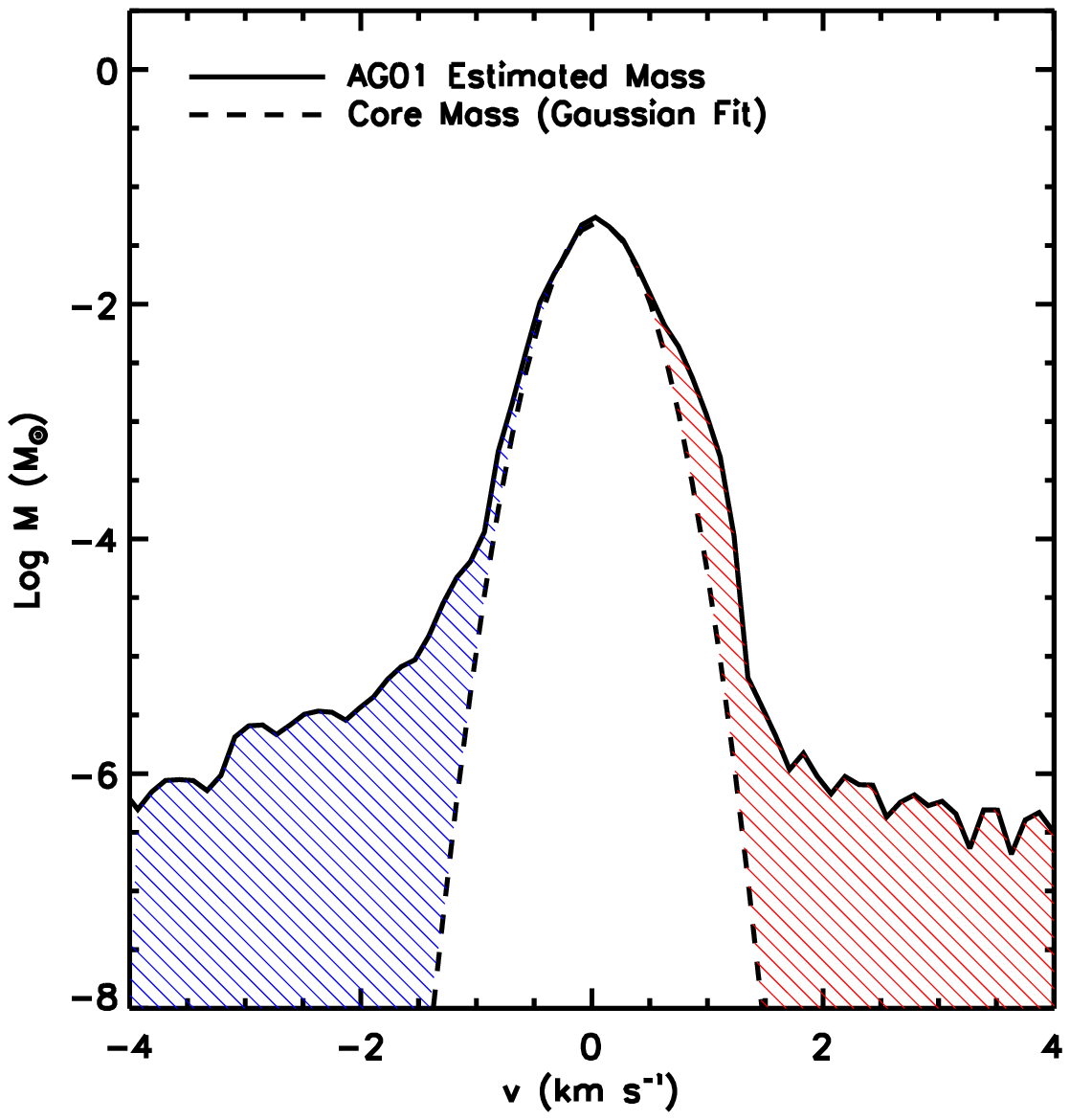}
%\plotone{pltgbw08130_4_lng_0.00000_lat_45.0000Massvfig.eps}
\caption{R1 (top) and R2 (bottom) gas mass versus velocity derived using the AG01 method. 
A Gaussian fit over the velocity range $-0.75~\kms \le v
  \le 0.75~\kms$ (dashed line) represents the non-outflow
  material. The shaded region shows the blue-shifted and red-shifted
  outflow mass. See AG01 Figure 7c for comparison. 
\label{mass_vs_vel} }
\end{figure}

An additional step is necessary to distinguish between cloud mass and
outflow mass. Summing over the total area, we obtain $M(v)$ as shown in Figure
\ref{mass_vs_vel}. AG01 found that the peak, which  includes the bulk of the core mass, was well fit by a
Gaussian. We find that a Gaussian provides a satisfactory fit in the case of R1, where there
is only a single outflow, while the R2 data, in which there is a
second ``hidden'' outflow, a Gaussian does not well describe all of the high
density gas (see Figure \ref{mass_vs_vel}). 
%over the bulk of the ambient cloud velocities ($\pm 0.75~\kms$). 
We then subtract this fit from $M(v)$ and integrate over all
velocities to get the outflow mass, i.e., the shaded area shown on Figure
\ref{mass_vs_vel}. 

\begin{figure}
\epsscale{2.4}
%\plotone{Mass_vs_tlat_90_vshift_4vw.eps}
%SSRO Note current R2 is 1'' need to switch to 4''
\plottwo{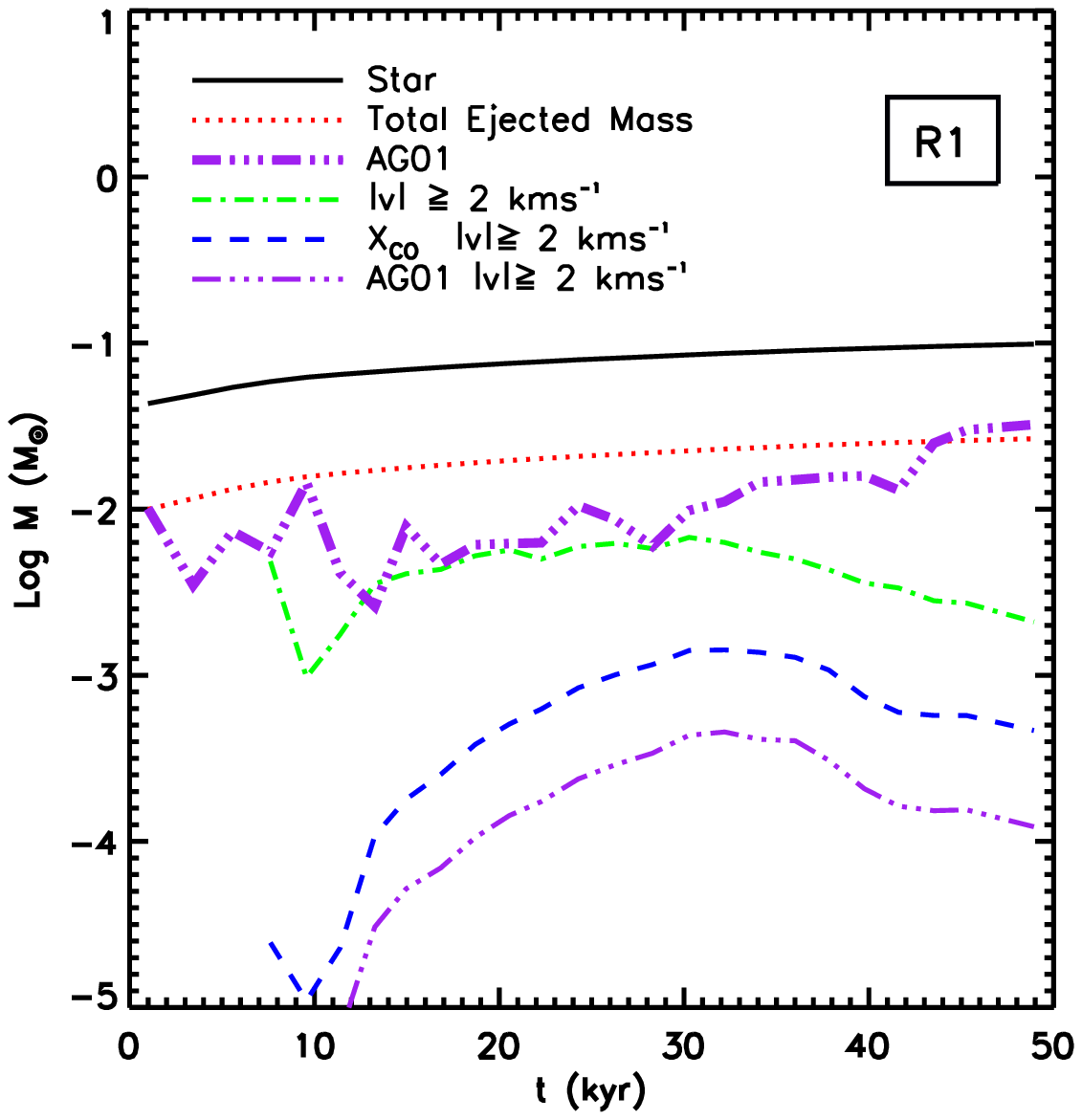}{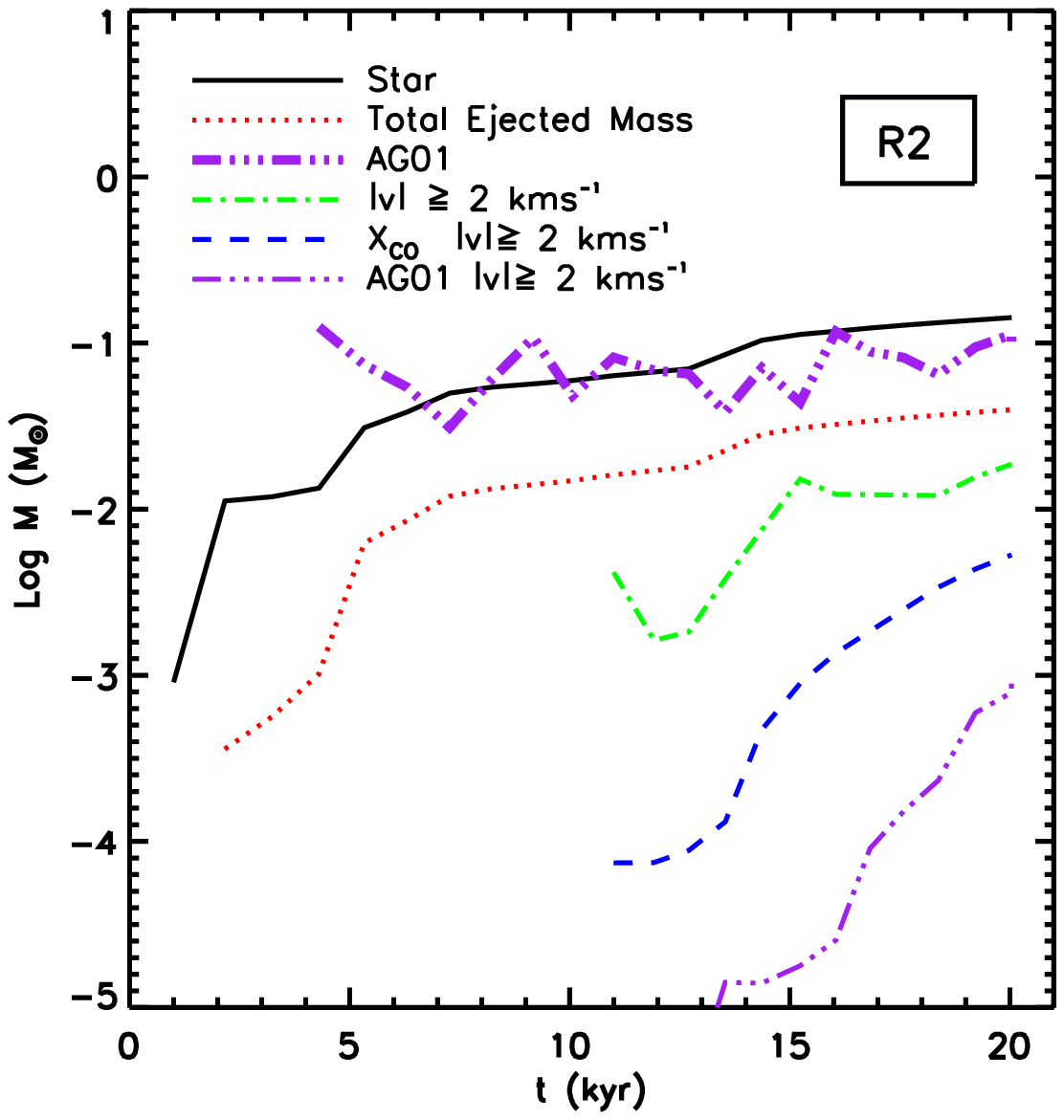}
\caption{ Star and outflow gas mass as a function of
  time for the outflows inclined approximately 45 degrees with respect
  to the line-of-sight. The total ejected mass (red dotted line) includes all mass ejected over all velocities in the
  outflow. The outflow mass (green dot-dashed line) is estimated from the raw
  simulation data for 
  $|v|\ge 2 \kms$.  The $X_{\rm CO}$ mass (blue dashed line) is estimated from the $^{12}$CO(1-0) emission observed with a 4''
  resolution beam in channels with $|v| \ge 2 ~\kms$. The thick purple
 dot dot-dot-dashed line shows
  the mass estimated via the AG01 method where a fit to the
  core gas mass is subtracted (see Figure \ref{mass_vs_vel}). The
  mass estimated from the AG01 method  where a simple velocity cut of
  2 $\kms$ is
  imposed in lieu of a fit is also shown (thin purple dot-dot-dot-dashed line).
Note that the R2 mass estimates include  
  both stars and both outflows.
%,
%  where the outflow masses are calculated in the frame of the most
%  massive star.
\label{synth_sim_mass} }
\end{figure}

Figure \ref{synth_sim_mass} shows the simulated outflow mass 
 compared with that derived using the $X$-factor and the
 AG01 method. For R2,
 Figure \ref{synth_sim_mass} displays the the total mass 
ejected by both stars. Although the
 axes of the two R2 outflows are not aligned, projection, beam
 resolution, close proximity, and the low-velocities of the outflows
 would prohibit
 individual identification of the two outflows observationally.

 In the figure, the ``total ejected mass'' is the total mass launched by the
 protostar  
 according to the outflow model. This is an essentially numerical quantity
 representing the total mass deposited in the zones around the protostar as
 described in Section 2. This quantity is distinct from what an
 observer would identify as the ``outflow mass,'' which is defined on the
 basis of velocity and thus also includes entrained gas.
 At any given time,  the \textit{launched} mass includes gas that
 goes on to comprise a
 low-velocity, less collimated component of the outflow and gas that
  mixes with the ambient gas becoming indistinguishable
 even with full information.\footnote{Some code methodologies, such as smooth particle hydrodynamics, permit the user to track individual fluid elements, but ORION does not have this capability.}
%SSRO
Since the launched material is placed on the grid near the protostar
 by design, this total does not include entrained 
 envelope gas. Once the wind is deposited in
 the grid it is impossible to differentiate launched gas from
 entrained, accelerated gas. In principal, due to the addition of
 entrained gas the total mass of high-velocity gas, i.e., the
 \textit{outflow} mass, could exceed the mass numerically ejected by the protostars.

For both runs, 
the mass derived from the raw simulation
 data with $|v|\ge 2~\kms$ and outflow orientation of 45 degrees is
 significantly less than the total ejected mass by a factor of 5-10. 
This reinforces the point that the majority of outflow gas in an
inclined outflow is simply not
 observable even with a perfect method for converting between CO emission
 and total gas mass. However, the integrated ejected outflow
 mass does not include gas that is entrained and accelerated by the
 outflowing material, so in principle, the simulated and observed gas could be
 larger than the total ejected mass.

The gas mass estimated from the CO emission using the $X$-factor
appears to track the mass of the high-velocity gas fairly well, although it is generally a factor of $5-10$
lower. This discrepancy is due to the high mean optical depth  of the
core ($\tau > 10$ for $^{12}$CO(1-0) and $\tau \sim$ a few for  $^{13}$CO(1-0)). 
%Simulation R2, which has a lower optical depth, exhibits less
% discrepancy between the $X_{\rm CO}$ derived mass and the actual mass. 
The AG01 method, which subtracts off the core mass, appears to perform somewhat better for most simulation
outputs. However, the accuracy of the mass estimation strongly
depends upon how well $M(v)$ can be fit by a simple function. 
For example,  at late times
the R1 $M(v)$ is slightly asymmetric, leading to a negative derived
outflow mass. Note that Figure \ref{mass_vs_vel} shows the log of the mass,
whereas the fit is dominated by a few points around $v\simeq0$ km
s$^{-1}$. 
Differences between the gas mass and fit at these low speeds far exceed the total mass contributed by
high-velocity gas in the distribution wings. To account for this, we only
subtract the curves beyond the full-width half-maximum of the Gaussian
fit and omit velocities where the difference is negative.
%Rewrite?
However, since the mass in the core is so much higher
than the mass in the high-velocity channels, the mass estimates 
strongly reflect the symmetry and Gaussianity of the mass
distribution. 
%Figure \ref{mass_vs_vel} shows that the
%``core'' could be fairly non-Gaussian.
The sucess
of AG01 in estimating the mass of outflow HH300 relies upon the
goodness of fit of a Gaussian and the relatively larger outflow
mass relative to the core mass.

If the outflow mass is instead defined 
as a sum over $M(v)$ where $|v| \ge 2~\kms$ then the mass estimate
follows a similar trend to the mass derived using the $X$-factor. 
However, this suggests that
the more complicated AG01 procedure still underestimates the total
outflow mass in the optically thick case.
The factor of $\sim$5
disrepancy between the two methods is due to different
assumptions about CO abundance.
%rather than a pronouncement on $X_{\rm CO}$. 

%Odyssey: _Outflows/idlcodes/plot_angle_time_evol.pro

\subsection{Outflow Temperature}

Observationally, $^{12}$CO(1-0) and $^{12}$CO(2-1) may be used in
combination to derive the excitation temperature, $T_{\rm
  ex}$. Temperature variation may be used to discriminate between
different outflow models \citep{arce02}. For example, shocked outflow gas
should be higher temperature than ambient gas, and the gas temperature
should rise with the maximum outflow velocity \citep{lee01}. 
For
optically thin gas, the excitation temperature may be related to the
ratio of the line intensities by 
\begin{equation}
R_{21/10} =4 e^{-11/T_{\rm ex}} \label{texcit},
\end{equation}
where $R_{21/10}$ is the ratio of the $^{12}$CO(2-1) to $^{12}$CO(1-0)
lines \citep{arce02}. 
%For the simulated outflows, both these transitions are
%optically thick.
Observers typically assume that the high-velocity gas
that constitutes the outflows is optically thin. We find that in many
cases line ratios in the
channels  with $|v| \ge 2 ~\kms$ are above 1.0, the line ratio limit in
the optically thick case, indicating that the gas is at most 
marginally optically thick in the outflow.

\begin{figure}
\epsscale{1.2}
\plotone{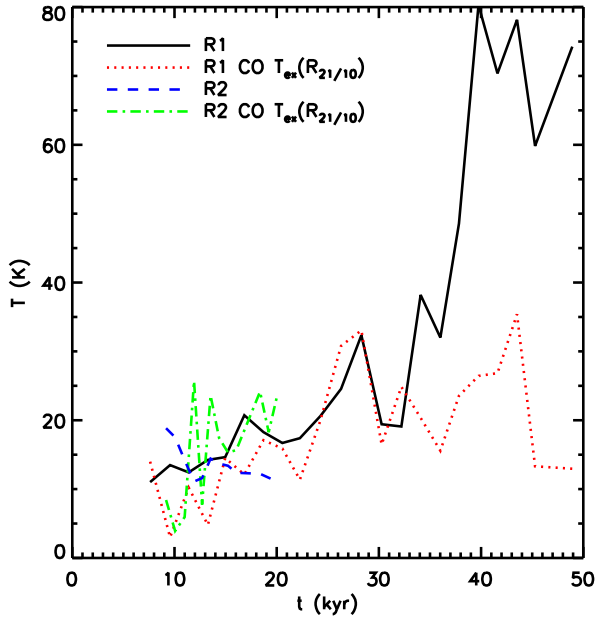}
\caption{ Outflow gas temperature versus time when the outflow is inclined
  45 degrees with respect to the line-of-sight. The R1 and R2 gas
  temperatures are computed as an average over the cells with
  velocities $|v_{45}|\ge 2 \kms$. The CO excitation temperature,
  $T_{\rm ex}$, is calculated using equation \ref{texcit} for an
  observation with 4'' resolution.
\label{synth_sim_temp} }
\end{figure}

% so equation
%\ref{texcit} should not be be an accurate approximation for $T_{\rm ex}$.
%For optically thick gas, the excitation temperature may be related to
%the ratio of the line intensities by
%\begin{equation}
%R_{21/10} =\frac{2(e^{5.5/T_{\rm ex}}-1)}{e^{11/T_{\rm ex}}-1} \label{texcit},
%\end{equation}
%where $R_{21/10}$ is the ratio of the $^{12}$CO(2-1) to $^{12}$CO(1-0)
%lines \citep{bachiller99}.
Figure \ref{synth_sim_temp} shows that the estimated excitation
temperatures are generally higher than the ambient gas temperature of 10
K. The excitation temperatures also increase with time, which is
expected if the shock strength and, hence post-shock temperature, increase
with time. This effect also occurs in the simulations, although a
rising gas temperature also results from increased protostellar
heating (see OKMK09). 
The two causes are difficult to distinguish from a single
average temperature value.
However, $T_{\rm ex}$ does not appear to be well correlated
with the actual gas temperatures in the simulation. 
The disagreement at later times is likely due to the increasing
optical depth of the
high-velocity gas.  \citet{hatchell99} show that gas with
$\tau=1$  and inferred $T_{\rm ex} = 30$ K under the assumption that the gas
is optically thin will
actually have $T_{\rm ex} = 40$K. The optically thin
approximation increasely underestimates temperatures for
higher optical depths and gas temperatures.

%Van Kempen Outflow
%http://adsabs.harvard.edu/abs/2009A%26A...501..633V

\section{Discussion}

Often only a single outflow lobe is observed \citep{arce10,
  ginsburg11}, creating uncertainty whether high gas velocites are
instead due to coherent turbulent motion. We
  find that even with completely symmetric bipolar energy injection,
  interactions between the outflow and the turbulent envelope can
  result in significant asymmetry between the lobes. Combined with
  inclination effects, it is therefore quite possible for observations
  to identify only one component. 

%SSRO, 12 out of 27 are single lobe, 44%
\citet{seale08}, who include 12 outflows with only a single visible lobe
in their photometrically identified sample,  attribute asymmetry to inclination
effects that 
increase obscuration of the far lobe by the core envelope. For outflows
identified by high-velocity, coherent motion, the
absence of a counterpart may be due to a combination of 
significant source advection, interference with the turbulent cloud velocities, or  asymmetric dynamical or magnetic interaction with the envelope. %or aymmetric cloud turbulent velocities 
Since dense, star-forming gas and young protostars are observed 
to move subsonically relative to their envelopes and the host cloud \citep{kirk10,
  offner09a}, obscuration and envelope interaction are the most 
likely causes of non-detected lobes. 
%Cite outflow modeling paper pointed out by Ned

Outflow-envelope interactions are
  also largely responsible for the inferred outflow opening
  angle. Since the launching region and angle of
  the outflow remain fixed in our simulations, the significant angle
  broading is a result of the outflow sweeping up sucessively larger
  solid angles of the core envelope. 
%SSRO
Over the course of the simulation, the mean density of the core does
not decrease 
significantly. However, the outflow mass and momentum, which are
coupled to the protostellar mass, increase with
time. 
Once the outflow breaks out of the dense envelope,  
%on the edge of the cavity 
an increasing amount of gas on the edge of the cavity is swept up, broadening the cavity.     
Thus, current observational
  resolution is likely not yet probing the outflow launching
  region.
Our hydrodynamic simulations are also 
  in surprisingly good agreement with opening angles inferred from observations, suggesting
  that the details of the magnetic field evolution may play a sub-dominant role on these scales.
 Consequently, evolving opening angle trends can not yet be used to
  distinguish between different theoretical models and instead reflect
  a generic characteristic of outflow-envelope interaction.

%The total momentum injected by the outflows
Exactly how outflows are responsible for driving turbulent
 motions in molecular clouds remains an open question in star
 formation.
Some simulations of $\sim$pc size clouds indicate that the momentum from
 outflows is sufficient to achieve quasi-steady state equipartition
 between turbulent and gravitational energy 
 \citep{nakamura07, wang10}. However, other simulations of jets indicate
that wellcollimated outflows drive supersonic turbulence very inefficiently, although they may contribute significant energy at lower velocities \citep{banerjee07}.
%However, outflows must not only account for
% the magnitude of the energy injection but the range of scales
% \citep{matzner}. If outflows extend only across several parces they
% cannot contribute driving motions with modes on the scale of the
% whole cloud. 
However, observations of some molecular clouds  suggest that the
 energy injected by outflows is insufficent by an order of magnitude or
 more than the observed cloud turbulence (\citealt{arce10,
 ginsburg11}, but see also \citealt{swift08,nakamura11}).
Such estimations include corrections for inclination effects and mass
 underestimations that amount to a factor of 4.
% After including the low-velocity gas, we find that mass estimations
% probing gas with $|v| \ge 2~\kms$ may be
% a factor of 3-10 too low. 
Previous
 studies have found that neglecting low-velocity material may result in
 underpredicting the outflow mass by at least a factor of 2
 \citep{margulis85}.
%This is quoted from Arce...
For
 outflows identified with minimum velocities of 10 $\kms$, 
the missing momentum may be even larger
 since outflows from the lowest mass sources and outflows aligned
 perpendicular to the line-of-sight may be missed altogether. 

The method of AG01, which includes lower velocity gas, exhibits the best
agreement with the launched mass. However, it is unclear how much of
the low velocity material, which dominates the mass estimate, is actually
non-Gaussian turbulent core gas. 
The underestimation of outflow mass by both CO methods that employ a
velocity cutoff underscores the
large uncertainties underlying these derivations. Such
techniques are likely more accurate at later stages when the core mass
has declined and more gas has been swept up by the
outflows. Nonetheless, this result has important implications for the
derivation of outflow momentum. In a worst case scenario, the masses
 of young, inclined outflows in optically thick regions are
 underestimated by factor of 50.

Despite the severity of the problem, an order of magnitude discrepancy relative
 to corrections already applied by observers is unlikely. In addition,
 the momentum contributed by outflows from
 very young low-mass stars is likely small compared to outflows from
 higher mass sources. Consequently, the discrepancy we demonstrate
 here is unlikely to account for the turbulent deficit measured in
 clouds $\ge 10$ pc. In addition,  
outflows must not only account for
the magnitude of the energy injection but the appropriate range of scales.
 \citep{matzner02, swift08}. If measured outflow extents range from $\sim$0.1-2
 pc then they will not contribute driving motions on
 larger scales.   Magnetohydrodynamic simulations of outflow driven
 turbulence confirm that the peak of the velocity power spectrum
 occurs at the maximum length scale of the driving outflows \citep{carroll09, carroll10}.
%See Arce et al. 2010
%Note--Wang et al. find that including magnetic fields increase the
% energy deposition of the outflow in pc sized clump, but it 

%\subsection{Variability}

\section{Conclusions}

We present results from self-gravitating, radiation-hydrodynamic
simulations of low-mass protostellar outflows using the adaptive mesh
refinement code, ORION. We  produce synthetic observations of the
simulations in $^{12}$CO and $^{13}$CO in order to make
direct comparisons with observations.

We formulate a quantitative prescription for measuring
outflow opening angles and test that it reproduces previous estimates
of observed outflow angles determined ``by eye.'' Using this method, we
find that both the simulation data  and synthetic observations of the
simulation data show opening angle evolution that is similar
to trends found by Arce \& Sargent (2006). However, we show beam resolution can
significantly broaden the observed outflow angle and may broaden
angles inferred by \citet{arce06} by as much as a factor of two.
Different inclinations with respect to the line-of-sight have a smaller effect on
the measured angles.

We find that the interaction between the
outflows and the turbulent core envelopes produces significant
asymmetry and variation in outflow properties. Outflowing gas running
into denser gas may be diverted or confined. This explains why
observations may sometimes only identify a single outflow lobe. 
The clumpy outflow geometetry in some
observed outflows may even result from outflow-envelope intereaction
rather than variable mass accretion/ejection.

Using CO isotopologues we estimate the observed gas masses in two ways.
First, we infer mass using $X_{\rm CO}$, the standard factor often used by
extragalactic observers to convert between CO emission and molecular gas mass.
%and
%excitation temperatures from $^{12}$CO(1-0) emission. We  apply
Second, we apply the method of \citet{arce01}, who combine
$^{12}$CO(1-0) and $^{13}$CO(1-0) data to obtain mass estimates.
We find that although the mass of the high-velocity outflow component tracks the actual
outflow mass, this gas is not a good estimate for the total integrated
outflow mass. Even with a perfect conversion between CO emission and
gas mass, outflow mass inferred only using emission from the high
velocity component underestimates the mass launched by a factor of
$\sim$ 2-5. 

Masses derived from the CO emission, which is a poor tracer of gas mass in the
optically thick limit, underestimate the actual outflow mass by an additional factor
of 5-10. Ideally, as \citet{arce01} suggest, lower-velocity outflow gas can be included in the outflow
estimate if the core mass can be modeled and subtracted from the emission.
However, in the simulations the core mass distribution does not necesarily have a symmetric distribution of gas
velocities, and so it is not well modeled with a simple functional
form. Consequently, the more complicated method of \citet{arce01}
produces a larger but not necessarily more accurate outflow mass.
%estimate than simply applying the
%standard $X$-factor in this case. 

%The underestimation of the outflow mass in both cases underscores the
%large uncertainties underlying derivations of outflow masses. Such
%techniques are likely more accurate at later stages when the core mass
%has declined and more gas has been swept up by the
%outflows. Nonetheless, this result has important implications for the
%derivation of outflow momentum. Rather than factors of 2 for
%inclination effects, \citet{} 

We find that the excitation temperatures derived from $^{12}$CO are not
necessarily well correlated with the simulation gas
temperatures. However, 
they indicate that the
high-velocity outflow gas is hotter than the ambient gas and that the
gas 
temperature increases with time, which is consistent with the raw data
from the simulations.
%at early times the two agree
%within a factor of two, while the two diverge as the simulated outflow
%gas becomes warmer.

%Since the outflow gas slows and mixes with the ambient
%material fairly rapidly, the net ejected mass may be a factor of 25
%larger than the outflow gas.
%We find that
%synthetic observations agree within a factor of two of the actual
%simulation properties, which is encouraging for observational
%analysis. Although the masses of the high-velocity material agrees, we
%find that this gas is not a good estimate for the total integrated
%outflow mass. Since the outflow gas slows and mixes with the ambient
%material fairly rapidly, the net ejected mass may be a factor of 100
%larger than the outflow gas. The inferred ejected mass further varies
%by a factor of 10 depending upon the outflow inclination.

In our simulations, the outflow launching region remains fixed
throughout. Consquently,  we conclude that observations of opening angle
evolution do not probe the outflow launching mechanism and instead reflect the
evolving interaction between the outflow and core envelope. Higher resolution
observations in addition to simulations including the effects of magnetic fields
will be needed to accurately investigate the physics of this innermost region.

\acknowledgements{ We thank Eric Keto for assistance with MOLLIE and
  Andrew Cunningham for technical improvements to ORION. We also thank
  Tom Robitaille, Chris Beaumont, Eric Keto, Michelle Borkin and Ned Ladd for helpful discussions and the referee, Robi Banerjee for helpful comments.
This research has
been supported by the NSF through grants AST-0901055 (SSRO) and
  AST-0908159 (EJL). 
The simulations and data analysis were performed on the Odyssey
  supercomputing cluster at Harvard University.}

\bibliography{outflowbib.bib}

\begin{thebibliography}{49}
\expandafter\ifx\csname natexlab\endcsname\relax\def\natexlab#1{#1}\fi

\bibitem[{{Arce} {et~al.}(2010){Arce}, {Borkin}, {Goodman}, {Pineda}, \&
  {Halle}}]{arce10}
{Arce}, H.~G., {Borkin}, M.~A., {Goodman}, A.~A., {Pineda}, J.~E., \& {Halle},
  M.~W. 2010, \apj, 715, 1170

\bibitem[{{Arce} \& {Goodman}(2001)}]{arce01}
{Arce}, H.~G. \& {Goodman}, A.~A. 2001, \apjl, 551, L171

\bibitem[{{Arce} \& {Goodman}(2002)}]{arce02}
---. 2002, \apj, 575, 928

\bibitem[{{Arce} \& {Sargent}(2005)}]{arce05}
{Arce}, H.~G. \& {Sargent}, A.~I. 2005, \apj, 624, 232

\bibitem[{{Arce} \& {Sargent}(2006)}]{arce06}
---. 2006, \apj, 646, 1070

\bibitem[{{Arce} {et~al.}(2007){Arce}, {Shepherd}, {Gueth}, {Lee}, {Bachiller},
  {Rosen}, \& {Beuther}}]{arce07}
{Arce}, H.~G., {Shepherd}, D., {Gueth}, F., {Lee}, C., {Bachiller}, R.,
  {Rosen}, A., \& {Beuther}, H. 2007, Protostars and Planets V, 245

\bibitem[{{Bally} {et~al.}(1999){Bally}, {Reipurth}, {Lada}, \&
  {Billawala}}]{bally99}
{Bally}, J., {Reipurth}, B., {Lada}, C.~J., \& {Billawala}, Y. 1999, \aj, 117,
  410

\bibitem[{{Banerjee} {et~al.}(2007){Banerjee}, {Klessen}, \&
  {Fendt}}]{banerjee07}
{Banerjee}, R., {Klessen}, R.~S., \& {Fendt}, C. 2007, \apj, 668, 1028

\bibitem[{{Bontemps} {et~al.}(1996){Bontemps}, {Andre}, {Terebey}, \&
  {Cabrit}}]{bontemps96}
{Bontemps}, S., {Andre}, P., {Terebey}, S., \& {Cabrit}, S. 1996, \aap, 311,
  858

\bibitem[{{Carolan} {et~al.}(2009){Carolan}, {Khanzadyan}, {Redman},
  {Thompson}, {Jones}, {Cunningham}, {Loughnane}, {Bains}, \&
  {Keto}}]{carolan09}
{Carolan}, P.~B., {Khanzadyan}, T., {Redman}, M.~P., {Thompson}, M.~A.,
  {Jones}, P.~A., {Cunningham}, M.~R., {Loughnane}, R.~M., {Bains}, I., \&
  {Keto}, E. 2009, \mnras, 400, 78

\bibitem[{{Carolan} {et~al.}(2008){Carolan}, {Redman}, {Keto}, \&
  {Rawlings}}]{carolan08}
{Carolan}, P.~B., {Redman}, M.~P., {Keto}, E., \& {Rawlings}, J.~M.~C. 2008,
  \mnras, 383, 705

\bibitem[{{Carroll} {et~al.}(2010){Carroll}, {Frank}, \&
  {Blackman}}]{carroll10}
{Carroll}, J.~J., {Frank}, A., \& {Blackman}, E.~G. 2010, \apj, 722, 145

\bibitem[{{Carroll} {et~al.}(2009){Carroll}, {Frank}, {Blackman}, {Cunningham},
  \& {Quillen}}]{carroll09}
{Carroll}, J.~J., {Frank}, A., {Blackman}, E.~G., {Cunningham}, A.~J., \&
  {Quillen}, A.~C. 2009, \apj, 695, 1376

\bibitem[{{Cernicharo} \& {Reipurth}(1996)}]{cernicharo96}
{Cernicharo}, J. \& {Reipurth}, B. 1996, \apjl, 460, L57+

\bibitem[{{Chernin} \& {Masson}(1995)}]{chernin&masson95}
{Chernin}, L.~M. \& {Masson}, C.~R. 1995, \apj, 455, 182

\bibitem[{{Cunningham} {et~al.}(2011){Cunningham}, {Klein}, {Krumholz}, \&
  {McKee}}]{cunningham11}
{Cunningham}, A.~J., {Klein}, R.~I., {Krumholz}, M.~R., \& {McKee}, C.~F. 2011,
  ArXiv e-prints

\bibitem[{{Dame} {et~al.}(2001){Dame}, {Hartmann}, \& {Thaddeus}}]{dame01}
{Dame}, T.~M., {Hartmann}, D., \& {Thaddeus}, P. 2001, \apj, 547, 792

\bibitem[{{Frerking} {et~al.}(1982){Frerking}, {Langer}, \&
  {Wilson}}]{frerking82}
{Frerking}, M.~A., {Langer}, W.~D., \& {Wilson}, R.~W. 1982, \apj, 262, 590

\bibitem[{{Ginsburg} {et~al.}(2011){Ginsburg}, {Bally}, \&
  {Williams}}]{ginsburg11}
{Ginsburg}, A., {Bally}, J., \& {Williams}, J. 2011, in prep

\bibitem[{{Glover} \& {Mac Low}(2011)}]{glover11}
{Glover}, S.~C.~O. \& {Mac Low}, M. 2011, \mnras, 412, 337

\bibitem[{{Hatchell} {et~al.}(1999){Hatchell}, {Fuller}, \&
  {Ladd}}]{hatchell99}
{Hatchell}, J., {Fuller}, G.~A., \& {Ladd}, E.~F. 1999, \aap, 344, 687

\bibitem[{{Keto} \& {Rybicki}(2010)}]{keto10}
{Keto}, E. \& {Rybicki}, G. 2010, \apj, 716, 1315

\bibitem[{{Kirk} {et~al.}(2010){Kirk}, {Pineda}, {Johnstone}, \&
  {Goodman}}]{kirk10}
{Kirk}, H., {Pineda}, J.~E., {Johnstone}, D., \& {Goodman}, A. 2010, \apj, 723,
  457

\bibitem[{{Krumholz} {et~al.}(2004){Krumholz}, {McKee}, \&
  {Klein}}]{krumholz04}
{Krumholz}, M.~R., {McKee}, C.~F., \& {Klein}, R.~I. 2004, \apj, 611, 399

\bibitem[{{Ladd} {et~al.}(1998){Ladd}, {Fuller}, \& {Deane}}]{ladd98}
{Ladd}, E.~F., {Fuller}, G.~A., \& {Deane}, J.~R. 1998, \apj, 495, 871

\bibitem[{{Lee} {et~al.}(2000){Lee}, {Mundy}, {Reipurth}, {Ostriker}, \&
  {Stone}}]{lee00}
{Lee}, C., {Mundy}, L.~G., {Reipurth}, B., {Ostriker}, E.~C., \& {Stone}, J.~M.
  2000, \apj, 542, 925

\bibitem[{{Lee} {et~al.}(2001){Lee}, {Stone}, {Ostriker}, \& {Mundy}}]{lee01}
{Lee}, C., {Stone}, J.~M., {Ostriker}, E.~C., \& {Mundy}, L.~G. 2001, \apj,
  557, 429

\bibitem[{{Li} \& {Shu}(1996)}]{li&shu96}
{Li}, Z. \& {Shu}, F.~H. 1996, \apj, 468, 261

\bibitem[{{Margulis} \& {Lada}(1985)}]{margulis85}
{Margulis}, M. \& {Lada}, C.~J. 1985, \apj, 299, 925

\bibitem[{{Matzner}(2002)}]{matzner02}
{Matzner}, C.~D. 2002, \apj, 566, 302

\bibitem[{{Matzner} \& {McKee}(1999)}]{matzner99}
{Matzner}, C.~D. \& {McKee}, C.~F. 1999, \apjl, 526, L109

\bibitem[{{Matzner} \& {McKee}(2000)}]{matzner00}
---. 2000, \apj, 545, 364

\bibitem[{{Myers}(2009)}]{myers09}
{Myers}, P.~C. 2009, \apj, 706, 1341

\bibitem[{{Nakamura} {et~al.}(2011){Nakamura}, {Kamada}, {Kamazaki}, {Kawabe},
  {Kitamura}, {Shimajiri}, {Tsukagoshi}, {Tachihara}, {Akashi}, {Azegami},
  {Ikeda}, {Kurono}, {Li}, {Miura}, {Nishi}, \& {Umemoto}}]{nakamura11}
{Nakamura}, F., {Kamada}, Y., {Kamazaki}, T., {Kawabe}, R., {Kitamura}, Y.,
  {Shimajiri}, Y., {Tsukagoshi}, T., {Tachihara}, K., {Akashi}, T., {Azegami},
  K., {Ikeda}, N., {Kurono}, Y., {Li}, Z., {Miura}, T., {Nishi}, R., \&
  {Umemoto}, T. 2011, \apj, 726, 46

\bibitem[{{Nakamura} \& {Li}(2007)}]{nakamura07}
{Nakamura}, F. \& {Li}, Z. 2007, \apj, 662, 395

\bibitem[{{Offner} {et~al.}(2009{\natexlab{a}}){Offner}, {Hansen}, \&
  {Krumholz}}]{offner09a}
{Offner}, S.~S.~R., {Hansen}, C.~E., \& {Krumholz}, M.~R. 2009{\natexlab{a}},
  \apjl, 704, L124

\bibitem[{{Offner} {et~al.}(2009{\natexlab{b}}){Offner}, {Klein}, {McKee}, \&
  {Krumholz}}]{Offner09}
{Offner}, S.~S.~R., {Klein}, R.~I., {McKee}, C.~F., \& {Krumholz}, M.~R.
  2009{\natexlab{b}}, \apj, 703, 131

\bibitem[{{Pelletier} \& {Pudritz}(1992)}]{pelletier92}
{Pelletier}, G. \& {Pudritz}, R.~E. 1992, \apj, 394, 117

\bibitem[{{Pineda} {et~al.}(2008){Pineda}, {Caselli}, \& {Goodman}}]{pineda08}
{Pineda}, J.~E., {Caselli}, P., \& {Goodman}, A.~A. 2008, \apj, 679, 481

\bibitem[{{Pollack} {et~al.}(1994){Pollack}, {Hollenbach}, {Beckwith},
  {Simonelli}, {Roush}, \& {Fong}}]{pollack94}
{Pollack}, J.~B., {Hollenbach}, D., {Beckwith}, S., {Simonelli}, D.~P.,
  {Roush}, T., \& {Fong}, W. 1994, \apj, 421, 615

\bibitem[{{Robitaille} {et~al.}(2006){Robitaille}, {Whitney}, {Indebetouw},
  {Wood}, \& {Denzmore}}]{robit06}
{Robitaille}, T.~P., {Whitney}, B.~A., {Indebetouw}, R., {Wood}, K., \&
  {Denzmore}, P. 2006, \apjs, 167, 256

\bibitem[{{Rosen} \& {Smith}(2004)}]{rosen04}
{Rosen}, A. \& {Smith}, M.~D. 2004, \mnras, 347, 1097

\bibitem[{{Seale} \& {Looney}(2008)}]{seale08}
{Seale}, J.~P. \& {Looney}, L.~W. 2008, \apj, 675, 427

\bibitem[{{Semenov} {et~al.}(2003){Semenov}, {Henning}, {Helling}, {Ilgner}, \&
  {Sedlmayr}}]{semenov03}
{Semenov}, D., {Henning}, T., {Helling}, C., {Ilgner}, M., \& {Sedlmayr}, E.
  2003, \aap, 410, 611

\bibitem[{{Shu} {et~al.}(1994){Shu}, {Najita}, {Ruden}, \& {Lizano}}]{shu94}
{Shu}, F.~H., {Najita}, J., {Ruden}, S.~P., \& {Lizano}, S. 1994, \apj, 429,
  797

\bibitem[{{Stone} {et~al.}(1998){Stone}, {Ostriker}, \& {Gammie}}]{stone98}
{Stone}, J.~M., {Ostriker}, E.~C., \& {Gammie}, C.~F. 1998, \apjl, 508, L99

\bibitem[{{Swift} \& {Welch}(2008)}]{swift08}
{Swift}, J.~J. \& {Welch}, W.~J. 2008, \apjs, 174, 202

\bibitem[{{Wang} {et~al.}(2010){Wang}, {Li}, {Abel}, \& {Nakamura}}]{wang10}
{Wang}, P., {Li}, Z., {Abel}, T., \& {Nakamura}, F. 2010, \apj, 709, 27

\bibitem[{{Yu} {et~al.}(1999){Yu}, {Billawala}, \& {Bally}}]{yu99}
{Yu}, K.~C., {Billawala}, Y., \& {Bally}, J. 1999, \aj, 118, 2940

\end{thebibliography}
\bibliographystyle{apj}

\end{document}